\documentclass{article}

\usepackage{graphicx}

\begin{document}
 
\begin{center}
{\Large \bf
Neutrinos} 
\\[0.5cm]
{\large \bf 
S.M. Bilenky} 
\\
{\it
{
Joint Institute for Nuclear Research, Dubna, Russia, \\
and  
\\ 
Physik Department, Technische Universit{\"a}t M{\"u}nchen,
\\
James-Franck-Strasse, D-85748 Garching bei M{\"u}nchen, Germany
}}

\end{center}

\begin{abstract}
The general status of neutrino physics are given. The history of
the neutrino, starting from Pauli and Fermi, is presented. The phenomenological
V-A theory of the weak interaction and the
unified theory of the weak and electromagnetic interactions, the so-called
Standard Model, are discussed. The problems of 
of neutrino masses, neutrino mixing, and neutrino oscillations are
discussed in some details.
\end{abstract}

\tableofcontents

\vskip5mm

\section {Introduction}

Neutrinos are elementary particles. Three flavor neutrinos
exist in
nature:
the electron neutrino $\nu_e$, the muon neutrino $\nu_{\mu}$ and the tau
neutrino
$\nu_{\tau}$.

Neutrinos are members of the three lepton families. 
Other particles that are members of the  families are,
correspondingly,the
electron $e^-$, the muon $\mu^-$ and the tau $\tau^-$. There are also
three
families of other elementary particles, the quarks:  
($u$, $d$), ($c$ , $s$)  and ($t$, $b$)

There are three fundamental interactions of elementary particles
that are characterized by the strength of the interaction:
strong, electromagnetic and weak. There is also the fourth gravitational
interaction between
particles. However, it is so weak that it can be neglected at all
available
energies.

The strong interaction is the
interaction between quarks and gluons, neutral particles with spin 1.
The interaction between quarks is the result of the exchange of
gluons.
Protons, neutrons, pions and all other hadrons are bound states of
quarks.

The electromagnetic interaction is the interaction between charged
particles and $\gamma$ -quanta.
The Coulomb interaction between charged particles is due to the exchange
of 
photons.
Atoms of different elements are bound states of electrons and nuclei. 

The weak interaction is the interaction between 
fundamental fermions (quarks, charged leptons, neutrinos) 
and
charged $W^{\pm}$ and neutral $Z^0$ bosons, heavy particles with spin 1 .
For example, the $\beta$-decay of the neutron
 \begin{equation}
           n \to p + e^{-} + \bar \nu
 \end{equation}
is due to the exchange of a charged $W$- boson between
$e -\nu$ and $d-u$ pair in nucleons.
Because of weak and electromagnetic interactions, all
particles ,except the electron, proton and neutrinos are
unstable. For example, the $\pi^+$-meson decays
into $\mu^+$ and $\nu_{\mu}$. The muon $\mu^+$ decays into $e^+$ ,
$\bar\nu_{\mu}$ and $\nu_e$ and so on. It was established during 
the last thirty years that the weak and electromagnetic
interactions are 
parts of a 
{\it electroweak interaction} 

Quarks take part in strong, electromagnetic and weak interactions.
Char\-ged
leptons - in electromagnetic
and 
weak interactions. Neutrinos are exceptional elementary particles:
their electric charge is equal to zero and they 
take part only in the weak
interaction.   
The role of neutrinos in physics and astrophysics is
determined by
this fact. 

The investigation of neutrino processes allows to one
obtain important information on the structure of the weak interaction.
The detailed study 
of the scattering of high energy neutrinos on nucleons
was very important for the establishment of 
quark structure of the
nucleons.
The detection of solar neutrinos  
allows one to investigate
the internal invisible region of the sun, where solar energy is
produced etc.

Neutrinos are also exceptional particles because of their internal
properties.
The masses of the neutrinos are much smaller than the masses of the
corresponding
family partners. 
Because of small neutrino masses and the so called neutrino mixing new
neutrino processes {\it neutrino
oscillations }, periodical transitions between different flavor
neutrinos, become possible. Recently evidence in favor of
neutrino
oscillations was found in the Super-Kamiokande experiment in Japan.
This discovery and also the discovery 
of the deficit of solar neutrinos by the Homestake
and other solar neutrino
experiments opened a new field of research in neutrino physics:
the physics of massive and mixed neutrinos.
It is a general belief that neutrino
masses and neutrino mixing angles 
are
determined by new physics at a mass scale that is much larger than
the 
scale of the present-day physics (hundreds of GeV).

We will list here the most important discoveries, connected with
neutrinos.

\begin{enumerate}

\item In 1954-56 in the experiment of F. Reines and C.W. Cowan the
electron
neutrino
was discovered. For this discovery F. Reines was rewarded
by
the Nobel prize in 1994.

\item In 1956 in the experiment of 
C.S. Wu et al 
the parity violation  in $\beta $ -decay
was discovered.

\item In 1958 in the experiment of M. Goldhaber et al the helicity of
the neutrino
was measured and evidence for the left-handed two-component neutrino
was obtained.

\item In 1962 in the Brookhaven experiment the second type of neutrino, 
the muon neutrino, was discovered. In 1988 for this discovery L.Lederman,
J. Steinberger, and M. Schwartz were rewarded the Nobel prize
for this discovery.

\item In 1973 in experiments at the neutrino beam at CERN a new type of
weak
interaction, Neutral Currents, was discovered.

\item In the eighties in experiments on the measurement of deep
inelastic
scattering of neutrinos on nucleons the quark structure of nucleons was
revealed and established.

\item In 1970 in the experiment of R. Davis et al neutrinos from the sun
were
detected. In these experiments and also in the GALLEX,
SAGE,
Kamiokande and Super-Kamiokande solar neutrino experiments
the existence of a solar neutrino problem
(deficit of solar $\nu_e$'s)
 was discovered

\item In 1987, in the Kamiokande, IMB and Baksan experiments,  
neutrinos from the explosion of the Supernova SN1987A in the Large
Magellanic
Cloud were detected.

\item In 1998, in the Super-Kamiokande experiment, compelling
evidence
in favor of oscillations of atmospheric neutrinos
was found.
\item  $ \ldots $
\end{enumerate}

Here we present 
at an elementary level
only the basics of neutrinos in 
particle physics. Those, who like to study
this interesting and exciting field of physics must read the original
papers and
books. Some books and recent reviews are listed in the bibliography.

\section{The history of the neutrino. Pauli}                 

The history of the neutrino 
started in
1930 with the
proposal of W. Pauli.
At that time
the electron $e^-$ and proton
$p$ were considered as the only elementary particles.
It was assumed that the nuclei of all elements heavier than hydrogen
are
bound
states of electrons and protons. 

In the framework of this assumption there were two fundamental problems.
The first problem was connected with the spectrum of energies of
electrons in $\beta$-decay, the process of the decay of a nucleus with
emission
of an electron. If some nucleus $A$ is transferred into another nucleus
$A'$
with the emission of
an electron then, according to the law of the conservation of the energy
and
momentum, the energy of the electron must be approximately equal to 
$M_A -M_{A'}$ 
($M_A$ and $M_{A'}$ are masses of the initial and
final nucleus). 
However,
in experiments on the investigation of $\beta$-decay a continuous
spectrum of energies $E$ up to $E_{0} \simeq M_A -M_{A'}$  was
observed.

The second problem was the problem of the spin
of the nitrogen  $^7 N_{14}$ and  other nuclei. The
atomic
number of $^{7} N_{14} $
is equal to 14 and the charge of the nucleus is equal to 7$e$ .
If we assume that nuclei are bound states
of protons and
electrons, the $^7 N_{14}$ nucleus is a bound state of 14 protons and 7
electrons. The spins of the proton and electron
are equal to 1/2. 
Thus, for the spin of the
$^7 N_{14}$
nucleus we will obtain half-integer value. 
However, from
experiments on the investigation of the spectrum of
 $^7 N_{14}$ molecules
it was known that $^7 N_{14}$ nuclei satisfy Bose statistics and,
according to the theorem on the connection between spin and
statistics,
the spin of the $^7 N_{14}$ nucleus must be an integer. 
This problem was known as the ``nitrogen catastrophe".

In order to solve these problems Pauli assumed 
that there exists in nature
a neutral particle with spin 
1/2,  mass less than the electron mass and
with a mean free path
much larger than the mean free path of a photon.
Pauli called this particle "neutron"  and he assumed 
that not only $p$'s and $e$'s but also "neutron"'s are constituents of
nuclei.
This assumption allowed him to solve easily the problem of the spin of 
 nitrogen and other nuclei. In fact, if in the $^7 N_{14}$ nucleus there
are an odd
number of "neutrons"
the spin 
of
this nucleus will be an integer. 

In order to explain  $\beta$-spectra,
Pauli assumed that in the process of $\beta$-decay the electron is emitted
together with
a "neutron" which is not detected in an experiment because of its large
mean
free path.
The energy released in $\beta$ -decay is shared between the electron and
"neutron" and as a result the continuous spectrum of energies of electrons
will be
observed. 

In 1932 the particle that today is called the neutron (the particle with
a
mass approximately equal to
the
mass
of the proton and the spin equal to 1/2)  was discovered by J. Chadwick
in the
nuclear
reaction

\begin{equation}
{}^{4}\rm{He} +{ }^{9}\rm{Be} \to { }^{12}\rm{C} +n 
\end{equation}

Soon after the discovery of the neutron, it was assumed independently
by W. Heisenberg, E. Majorana and D. Ivanenko 
that the
real constituents of nuclei are protons and neutrons.
This assumption allowed to explain all existing nuclear data.
In particular, according to this assumption the nucleus
 $^7 N_{14}$ is a bound state of 7 protons and 7 neutrons and the spin of
this nucleus must be an integer. Thus, the ``nitrogen
catastrophe" disappeared.

\section {The first theory of $\beta$ - decay. Fermi }

In 1933-34 E. Fermi proposed the first theory of the $\beta$-decay of
nuclei.
The Fermi theory was based 
on the assumption that nuclei are bound states of protons and neutrons
and
on the Pauli
hypothesis of the existence of
a
neutral, light, spin 1/2 particle with a large mean free path. Fermi
baptized 
this
particle with the name neutrino (from Italian neutral, small). 
Following Pauli, Fermi assumed  that in  
 $\beta$- decay the electron is emitted together with the neutrino.
The problem was to understand how an electron-neutrino pair is
emitted from a nucleus which is a bound state of protons
and neutrons.

For Fermi it was important an analogy with
electrodynamics. According to quantum electrodynamics 
in the transition of an electron  from 
an excited
state of an atom into a lower state a photon is emitted. 
In analogy with this process Fermi assumed that the
electron-neutrino pair {\it is produced in the process of the
quantum transition} of a
neutron inside a nucleus into a proton

\begin{equation}
n \to p + e^- +\nu
\label{BEN}
\end{equation}

The first theory of $\beta$- decay
was also built by Fermi
in  close analogy with quantum electrodynamics.
The main quantity of the quantum field theory is 
the density of the energy of the interaction, that is called 
the Hamiltonian of the interaction.

The Hamiltonian of the electromagnetic interaction 
has the form of the scalar product of the electromagnetic current
$j^{em}_{\alpha}(x)$
and the electromagnetic field $A^{\alpha}(x)$

\begin{equation}
{\cal{H}}_I^{\rm{em}}(x) = 
e \ j^{\rm{em}}_{\alpha}(x) \
A^{\alpha}(x)
\end{equation}

where the sum over $\alpha =0,1,2,3$  is assumed.
The electric charge $e$  characterizes the strength of the
electromagnetic interaction. 

The electromagnetic current
$j^{em}_{\alpha}$ is a 4-vector. The time component 
$j^{em}_0$ is the density of charge and the space components $j^{em}_i$
($i$=1,2,3) are components of the vector current.
The electromagnetic field $A^{\alpha}$ is also a 4-vector:
$A^0$ is a scalar potential and $A^i$ are components of a vector
potential.

 The electromagnetic
current of protons is given by

\begin{equation}
j^{\rm {em}}_{\alpha}(x) = \bar{p}(x) \gamma_{\alpha} p(x)
\end{equation}

Here $\gamma_{\alpha}$ are the Dirac matrices and $p(x)$ is the proton
field.

In analogy with (5)  Fermi assumed that the Hamiltonian
of 
$\beta$- decay had the form of the scalar product of the 
proton-neutron and electron-neutrino  currents

\begin{equation}
{\cal{H}}_{I}^{\beta}=
G_F (\bar{p} \gamma_{\alpha} n) (\bar{e} \gamma^{\alpha} \nu) \ + \
h.c.
\end{equation}

Here $G _F$ is the constant that characterize the strength of the
$\beta$ -decay interaction ( $G_F$ is
called
Fermi constant)$, n(x)$ is the field of neutrons, $e(x)$ is the field of
electrons
and $\nu (x)$ is the field of neutrino.

In quantum field theory $n(x)$ is the operator which annihilates the
neutron in
the initial state, the
operator $\bar p(x)$  creates the proton in the final state
and the operators  $\bar e(x)$ and $\nu (x)$ create the final electron
and
neutrino. 

The Fermi theory allows one to describe the 
$\beta$- decay of different nuclei.
This theory, however, could not describe all
$\beta$- decay data.
In 1936 Gamov and Teller generalized  the Fermi theory by including 
in the Hamiltonian
additional
scalar, tensor, pseudovector and pseudoscalar terms 
with four additional
interaction constants.

All $\beta$- decay data, existing at that time, could be described by
the Fermi-Gamov-Teller
interaction. This was an indirect evidence of the 
correctness of the Pauli-Fermi hypothesis of the neutrino.
The direct proof of the existence of the neutrino was obtained only in the
beginning of the fifties in the F. Reines and C.L. Cowan experiment.
We will discuss this experiment in the next section.
Let us start with a discussion of the
notion of {\it lepton number }.
   
\section{Lepton number. The Discovery of the neutrino }

As is well known, the total electric charge is conserved. This
means
that only such processes are allowed in which the sums of the electric
charges
of the initial and final particles are equal.   

According to quantum field theory
every charged particle has its {\it antiparticle}, a particle 
with the same mass and spin but opposite charge.
This general consequence of 
quantum field theory is confirmed by  all existing experimental
data. The antiparticle of the electron is the positron.
The electron and the positron have the same mass and the same spin and
the electric
charges of the electron and positron are equal to
$-e$ and $e$, respectively.
The existence of the positron was predicted on the basis of the Dirac
theory of the electron.
The positron was
discovered by C.D. Anderson in 1932. 
The antiparticle of the proton is the antiproton $\bar{p}$, a particle
with electric charge equal to $-e$ and a mass equal to the proton
mass. The antiproton was discovered in 1955 by O. Chamberlain, 
E.G. Segre et al. The antineutron $\bar{n}$ was discovered in 1956 and
so on.

Except electric charge there exist other conserved charges.
One such charge is the {\it baryon number}. The baryon numbers of 
$p$ and $ \bar{p} $ are equal to 1 and -1, respectively. The baryon
numbers of the $\pi ^{\pm}$ -mesons,  $ K ^{\pm}$,  $\gamma $-quantum
and other bosons
are equal to zero.   
Due to the conservation of the baryon number
the proton is a stable particle.

Let us now return to the neutrino. The fact that the neutrino is produced
in $\beta$-decay together with an electron suggests that there exist 
some conserved quantum number that characterizes these particles. 
This number is called lepton number.
Let us assume that the lepton numbers of the electron and the neutrino are
equal
to 1 and lepton numbers of the proton, neutron, photon and other
particles
are equal to zero. According to the general theorem, we mentioned before, 
the
lepton number of the positron is equal to -1 and the antineutrino, the
particle
with the lepton number equal to -1, must exist.   
{}From the conservation of lepton number it follows that
in $\beta$-decay together with an electron an {\it antineutrino} is
emitted.
We will discuss later the experiment 
in which evidence in favor of conservation of lepton number was obtained.

Now we will consider the experiment of F. Reines and C.L. Cowan 
in which the (anti)neutrino was discovered. 
In this experiment antineutrinos that are produced 
in $\beta$-decays of different nuclei, products of the fission of 
$U$ and $Pu$  in a reactor, were detected via the observation of the
process

\begin{equation} 
\bar{\nu} + p \to e^{+} + n
\label{NUP}
\end{equation}

A reactor is a very intense source of antineutrinos: about $2 \times 
10^{14}$ antineutrinos  per second 
are emitted per kW, generated by the reactor. The power of a modern  
reactor is about 4 GW. Thus, about $10^{21}$ antineutrinos are emitted by a
reactor per second. 
The experiment of Reines and Cowan was done at the Savannah
River reactor in USA. 
The detector in this experiment was a  liquid scintillator loaded
with cadmium.
The positron, produced in the process (7), quickly slowed down to 
rest and
annihilated with an electron into two 
$\gamma$-quanta with  energies 
about $E=m_e \simeq 0.51 MeV$, moving in opposite directions.
 These 
$\gamma$-quanta
were detected by photomultipliers connected with scintillators.

The neutron produced in the process (7) was slowed down and was captured
by
a cadmium nucleus emitting $\gamma$-quanta with total energy about 9MeV.
These $\gamma$-quanta 
give a several microseconds delayed signal in the photomultipliers.

The probability of interaction is characterized in physics by the {\it
cross
section} that has dimension of (length)${}^2$. 
In order to determine the cross section we will consider the
flux of the particles that pass through the matter.
Let us consider the element of the volume of the target 
with unit area oriented perpendicular to 
the momentum $\vec{p}$
(see Fig.~1 )
\begin{figure}[h]
\begin{center}
\includegraphics*{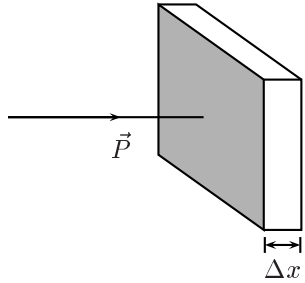}
\end{center}
\caption{The element of the volume of the target oriented perpendicularly
to the momentum $\vec p$.}
\label{fig1}
\end{figure}
 The number of particles of the target in this
volume is equal to $ 1 \cdot \Delta x \cdot \rho $ 
($\rho$ is the number density of the target).
The cross section $ \sigma $ of a process of scattering, absorption,... 
is the probability of the process
per one particle in the
target and per unit flux. For the change of the flux  
after passing
trough the
element, shown in Fig.~1 we have
 
\begin{equation}
\Delta I(x) =
I(x+\Delta x) - I(x) =
-\rho \sigma \Delta x I(x)
\label{DEL}
\end{equation}

{}From (\ref{DEL})we obtain 

\begin{equation}
I(x) = e^{-\rho \sigma x} I(0)
\label{IX}
\end{equation}

where $x$ is the distance that the particles pass in the matter.
We can rewrite (\ref{IX}) in the form

\begin{equation}
I(x) = e^{-\frac{x}{L}}I(0)
\end{equation}

where $ L = \frac{1}{\rho \sigma}$ is the mean free path.

For the cross section of the process (\ref{NUP}) in the experiment of
Reines 
and
Cowan the following value was found
\begin{equation} 
 \sigma = (11 \pm 
4) \times{10^{-44}} ~\rm {cm^2}
\end{equation}

This is a very small cross section. 
Let us consider the propagation of reactor antineutrinos 
with an energy of a few MeV
in the earth.
We have  
$\sigma \simeq 10^{-43} ~\rm {cm^2}$
and $\rho \simeq 10^{24}$ protons per $\rm {cm^3}$. Thus, for the mean
free
path of a neutrino
in the earth we have
$L  \simeq 10^{14}$ km. Remember that the earth's diameter is about
$10^4$ km.
Thus, the probability for an antineutrino with an energy of a few MeV 
to interact with the matter of the earth is about $10^{-10}$!

The fact that the neutrino and the antineutrino are different
particles was established in the reactor experiment of R. Davis in 1955. 
As we discussed earlier, a reactor is a source of {\it antineutrinos}. 
If the lepton number is conserved, the reaction
\begin{equation} 
 \bar{\nu} + {}^{37}\rm{Cl} \to e^{-} + {}^{37}\rm{Ar} 
\label{CLAR}
\end{equation}
is forbidden. 
In the Davis experiment a large tank with carbon tetrachloride
($C_{2}Cl_{4}$)
liquid 
was irradiated over a long period of time by antineutrinos from
the reactor. 
After every run atoms of ${}^{37} \rm {Ar}$ were
extracted
from the
liquid by purging it with $^{4}He $ gas and they were put into 
a
low-background Geiger
counter. The
$\gamma$-quanta 
produced in the $e^-$ capture by $^{37}Ar$ were detected.
No effect was observed. For the cross section of the process
(\ref{CLAR})
it was found that
\begin{equation} 
 \sigma = (0.1 \pm 0.6) \times {10^{-45}} ~\rm {cm^2}
\end{equation}
If the neutrino and the antineutrino had been identical, for the cross
section of the
process
(\ref{CLAR}) the following value
\begin{equation} 
 \sigma = 2 \times {10^{-45}}~ \rm {cm^2}
\end{equation}
would have been expected.

\section {Nonconservation of parity in $\beta$-decay. The two-component
neutrino}

In 1956 in an experiment by C.S. Wu et al nonconservation of parity in
$\beta$-decay
was discovered.
This was a very important discovery in particle physics 
that drastically changed 
our understanding of
the weak interaction and the neutrino. 

In order to explain the phenomenon of parity
violation we must remember that there are two types of vectors
: (true) vectors
and
pseudovectors. The direction of a vector does not depend on the choice of
the
coordinate system. The direction of a pseudovector 
is changed if we change the handedness of the coordinate system.
Typical vectors are momentum, coordinate, electric field etc. Angular
momentum,
spin, magnetic
field etc are
pseudovectors. 

Let us consider two coordinate systems: some right-handed
system
and a
system with all axes directed
opposite to the direction of the axes of the first system.
The second system is left-handed one.
If some vector $ \vec{A}$ has components $A_i$ (i=1,2,3) in the first
system,
in the second system the coordinates of this vector will be $- A_i$.
If  $\vec{B}$ is pseudovector with coordinates $B_i$ in the first
system,
then in the second system its
coordinates will be
$B_i$ (pseudovector changes direction).
The transformation from the first system to the second one is called
inversion
or parity
transformation.

In the Wu et al experiment the $\beta$-decay of polarized nuclei
$^{60}Co$ nuclei was investigated. 
The polarization (the average value of the
spin) is a pseudovector. Let us consider in a right-handed system the
emission of an electron at an angle $\theta$ between the direction of
the polarization of the nucleus and the electron momentum (see Fig.~2).
\begin{figure}[h]
\begin{center}
\includegraphics*{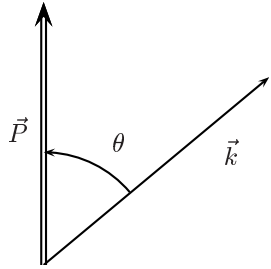}
\end{center}
\caption{The emission of an electron with the momentum $\vec k$
by a nucleus with polarization $\vec P$ (right-handed system).}
\label{fig2}
\end{figure}
In the left-handed system the direction of the polarization is
reversed and Fig.~2 corresponds to the emission of an electron at an
angle 
$\pi$ - $\theta$ (see Fig.~3).
\begin{figure}[h]
\begin{center}
\includegraphics*{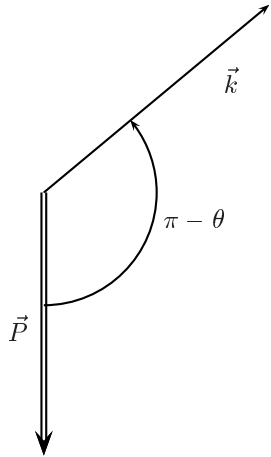}
\end{center}
\caption{The same as in Fig.~2 but in the
left-handed system.}
\label{fig3}
\end{figure}
The emission of an electron at the angle
$\pi$ - $\theta$ in the right-handed system corresponds to the emission of
an electron at the angle $\theta$ in the left-handed system. 
Thus,
right-handed
and left-handed systems are equivalent (the parity is conserved) if  
the number of electrons emitted (in a fixed system) at the angles $\theta$
and $\pi$ -
$\theta$
are equal. 

In the experiment of Wu et al a  large asymmetry
of the emission of the electrons with respect to the polarization of
the nuclei
was discovered. It
was observed that electrons are emitted 
predominantly in the
direction opposite to the direction of the polarization of the nuclei.
Thus, it
was
proved that parity is not conserved in $\beta$-decay (the left-handed
and right-handed systems are not equivalent). Later it was shown that
parity is not conserved in other weak processes.

Let us now consider in a right-handed system the emission 
of a left-handed neutrino $\nu_L$, a neutrino with the projection of the
spin
on the direction of
momentum (helicity) equal to -1 . In the
left-handed system the projection of the spin on the vector of
momentum of the neutrino will be equal to +1 (spin changes direction) .
Thus, if parity is conserved the probabilities of
emission of
the left-handed
neutrino
$\nu_L$ and the right-handed neutrino $\nu_R$ 
(in a fixed system)
must be the same:
\begin{equation} 
 w(\nu_L)=w(\nu_R)  
\label{VIOL}
\end{equation}
The discovery of the nonconservation of parity in weak interactions means 
that
these probabilities are not equal.

In 1957 Landau, Lee and Yang and Salam
proposed the theory of the {\it two
component neutrino}. This theory is based on the assumption that the mass
of the neutrino is equal to
zero. According to the theory of the two-component neutrino
for the neutrino there are only two possibilities:
\begin{enumerate}
\item 
the neutrino
is a left-handed particle $\nu_L$ and the antineutrino is a right-handed
antiparticle $\bar\nu_R$;
\item
the neutrino
is a right-handed particle $\nu_R$ and the antineutrino is a left-handed
antiparticle $\bar\nu_L$.
\end{enumerate}

In both cases the equality (\ref {VIOL}) is violated maximally.

The helicity of the neutrino was measured in 1957 in a spectacular
experiment by Goldhaber et al.
In this experiment neutrinos were produced in the
K-capture

\begin{eqnarray}
e^- + \rm {Gd} \to \nu_e + \null & \rm {Sm}^* &
\nonumber
\\
& \downarrow &
\nonumber 
\\
& \rm{Sm} & \null + \gamma
\end{eqnarray}

The measurement of the circular polarization of $\gamma$-quantum 
from the decay of $Sm^{*}$
allowed to
determine the helicity of the neutrino. The two-component neutrino theory
was
confirmed by this experiment and
it was established that the neutrino is a particle with negative helicity.
(see Fig.~4).
\begin{figure}[h]
\begin{center}
\includegraphics*{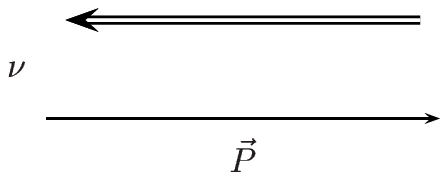}
\hspace*{1cm}
\includegraphics*{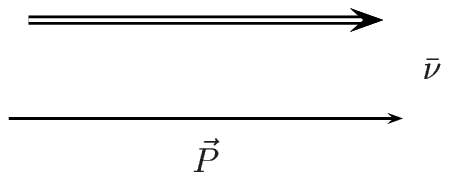}
\end{center}
\caption{Helicities of two-component neutrino and antineutrino. The
vector of the spin (momentum) of neutrino (antineutrino) is shown by
double line
(single line).}
\label{fig4}
\end{figure}

\section{Universal current $\times$ current theory of weak interactions}

The discovery of parity nonconservation in the weak interaction
and  the confirmation of the theory of a two-component neutrino led to 
an enormous
progress in the development of the weak interaction theory
(Feynman and Gell-Mann, Marshak and Sudarshan 1958).
At that time not only $\beta$-decay, but also other weak processes were
known. One such processes is $\mu$-capture
\begin{equation} 
 \mu^{-}+p \to \nu +n  
\label{CAPT}
\end{equation}

The first idea of a possible interaction, responsible for the decay
(\ref{CAPT}), 
was put forward by B.Pontecorvo. He compared the
probabilities
of $\mu$-capture and K-capture of an electron by a nucleus and came to the
conclusion that
the corresponding interaction constants are of the same order.
B. Pontecorvo assumed that there exists a {\it universal weak interaction}
that includes $e$ - $\nu$ and $\mu$- $\nu$ pairs. The idea of
 $\mu$ - $e$
universality was proposed also by G. Puppi, O.Klein and other
authors.

Let us notice that any fermion field $\psi(x)$ can be presented as a sum
of a
left-handed component
$\psi_{L}(x)$ and a right-handed component $\psi_{R}(x)$
\begin{equation} 
  \psi(x) =\psi_{L}(x) + \psi_{R}(x)  
\end{equation}
where
\begin{equation} 
 \psi_{L,R}(x) = \frac{1\mp\gamma_5}{2} \ \psi(x)  
\end{equation}
and $\gamma_5$ is a Dirac matrix.

The fact that the neutrino is a particle with negative helicity means that 
the field
of a neutrino
is a left-handed field $\nu_L$. 
Feynman and Gell-Mann, Marshak and Sudarshan assumed that
in the Hamiltonian of the weak interactions enter
{ \it left-handed components of all fields}. 
If we will make this assumption the Hamiltonian of $\beta$-decay
takes the very simple form

\begin{equation}
{\cal{H}}_I^{\beta}=
\frac{G_{\rm{F}}}{\sqrt{2}}~4~({\bar{p}}_L \gamma^{\alpha} \nu_L)
({\bar{e}}_L \gamma_{\alpha} \nu_{L}) \ + \ h.c.
\end{equation}

This interaction, like the Fermi interaction, is
characterized by only one interaction constant $G_F$.
It contains, however, parity-conserving 
(vector $\times$ vector and axial $\times$ axial) and 
parity-violating (vector $\times$ axial and axial $\times$ vector) parts.

Assuming $\mu - e$ universality, Feynman and Gell-Mann proposed the
theory that allowed one to describe all the weak processes known at that
time
and to predict new weak processes. They assumed that there exists a {\it
weak
current}

\begin{equation}
j^{\alpha}=2 \ [\bar{p}_L \gamma^{\alpha} n_L \ + \ 
\bar{\nu}_{eL} \gamma ^{\alpha} e_L \ + \
\bar{\nu}_{\mu L} \gamma^{\alpha} \mu_L]
\label{WECUR}
\end{equation}
and that the Hamiltonian of the weak interaction has the simple
current $\times$
current
form 

\begin{equation}
{\cal{H}}_I =
\frac{G_{\rm{F}}}{\sqrt{2}} \ j^{\alpha} \
j^{+}_{\alpha}
\label{HW}
\end{equation}
where

\begin{equation}
j_{\alpha}^+ = 2 \ [ \bar{n}_L \gamma_{\alpha} p_L \ + \
\bar{e}_L \gamma_{\alpha} \nu_{eL} \ + \
\bar{\mu}_L \gamma_{\alpha} \nu_{\mu L}]
\end{equation}
is the conjugated current.

In (\ref{WECUR}) the neutrino field that enters into the current
together
with
the electron
field (muon field) is denoted by $\nu_e$ ($\nu_{\mu}$). We will call
the corresponding
particles the electron neutrino and the muon neutrino. It was proved 
in the famous
1962
Brookhaven neutrino experiment
that $\nu_e$ and $\nu_{\mu}$ are different particles. We
will discuss this experiment in the next section.
Now we will continue the discussion of the
current $\times$ current Hamiltonian.
There are terms of two types in the Hamiltonian
(\ref{HW}): nondiagonal and
diagonal. Nondiagonal terms are given by

\begin{eqnarray}
{\cal{H}}_I^{\rm{nd}}=
\frac{G_{\rm{F}}}{\sqrt{2}}
4\{[(\bar{p}_L \gamma^{\alpha} n)( \bar{e}_L
\gamma_{\alpha}
\nu_{eL}) \ + \ h.c.] \ + \
\nonumber\\
+ \ [(\bar{p}_L \gamma^{\alpha} n_L)
(\bar{\mu}_L \gamma_{\alpha} \nu_{\mu L}) \ + \ h.c.] \ + \
\nonumber\\
+ \ [(\bar{e}_L \gamma^{\alpha} \nu_{eL}) 
(\bar{\nu}_{\mu L} \gamma_{\alpha} \mu_L) \ + \ h.c.]\}
\label{HND}
\end{eqnarray}

The first term of this expression is the Hamiltonian of $\beta$-decay  
of the neutron (\ref{BEN}), of the process

\begin{eqnarray}
\bar{\nu}_e + p \to e^+ + n
\end{eqnarray}
and other processes.

The second term of (\ref{HND}) is the Hamiltonian of
$\mu$-capture (\ref{CAPT}), of the process

\begin{equation}
  \nu_{\mu} + n \to \mu^{-} + p
\end{equation}

and other processes.

Finally the third term of (\ref{HND}) is the Hamiltonian of
$\mu$-decay

\begin{equation}
\mu^{+} \to e^{+} + \nu_e + \bar{\nu}_{\mu}
\end{equation}
and other processes.

Some processes that are described by
nondiagonal terms of the Hamiltonian
were observed in an experiment at the time when
the current $ \times $ current theory  was proposed.
 This theory also
predicted 
new weak processes such as
the process of elastic scattering of the electron antineutrino on
the electron

\begin{equation}
\bar{\nu}_e + e \to \bar{\nu}_e + e
\label{ANUEL}
\end{equation}
and others.
The Hamiltonian of these new processes is given by the diagonal terms of
(\ref{HW}):

\begin{equation}
{\cal{H}}^{\rm{d}}=
\frac{G_{\rm{F}}}{\sqrt{2}}
4[(\bar{\nu}_{eL} \gamma^{\alpha} e_L)
  (\bar{e}_L \gamma_{\alpha} \nu_{eL}) \ + \ldots]
\end{equation}

The predicted cross section of the process (\ref{ANUEL}) is
very small
and its measurement was a difficult problem. 
After many years of efforts 
F. Reines et al observed  
the process (\ref{ANUEL}) with
reactor antineutrinos. 

The
detailed investigation of this and another similar processes
showed that, except diagonal terms, in the Hamiltonian of
such processes there are additional  
neutral current (NC)
terms. We will discuss NC
later.

There were two alternatives for the weak interaction theory:
the current $\times{}$ current theory we described
and the theory with an intermediate vector charged $W^{\pm}$ - boson.
We will discuss now this last theory.
Let us
assume
that there exists heavy particles $W^{\pm}$ with spin equal to 1 and 
charges $ \pm e$ and that the fundamental weak interaction has the form

\begin{equation}
{\cal{H}}=\frac{g}{2\sqrt{2}} j_{\alpha} W^{\alpha} \ + \ h.c.
\end{equation}
where $g$ is the interaction constant and the current $ j^\alpha $ is
given by the expression (\ref{WECUR}). It is possible to show that at
energies much less than the mass of the $W$-boson
$m_{W}$ for the processes with a virtual (intermediate)  $W$-boson
the current $ \times{} $ current theory and the theory with the $W$-boson
are equivalent.

In fact, let us consider $ \mu$-decay
\begin{equation} 
 \mu^{-} \to e^{-} + \bar \nu_e + \nu_{\mu}  
\label{MUDEC}
\end{equation}

In quantum field theory the processes are described by
Feynman diagrams that are the convenient language and computational tool
of physicists. 
With the help of special
rules Feynman diagrams allows one to
calculate the probabilities of decay, cross sections and other
measurable quantities.

In the current $\times{}$ current theory the decay (\ref{MUDEC})
is the process of the first order in perturbation theory in the
constant $G_F$
and its Feynman diagram is 
presented in
Fig.~5.
\begin{figure}
\begin{center}
\includegraphics*[bb = 240 570 135 685, height=5cm]{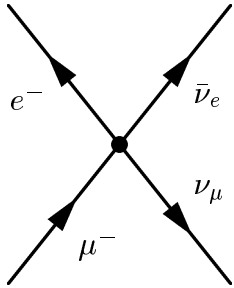}
\end{center}
\caption{The Feynman diagram of the decay $ \mu^- \to e^- {\bar {\nu}
_e } \nu_\mu $ in the current $\times$ current theory.}
\label{fig5}
\end{figure}
In the theory with  the $W$-boson the decay (\ref{MUDEC})
is the process of second order in perturbation theory in the
constant $g$. The Feynman diagram of the process is presented in 
Fig.~6.
\begin{figure}
\begin{center}
\includegraphics*[bb = 265 565 135 685, height=5cm]{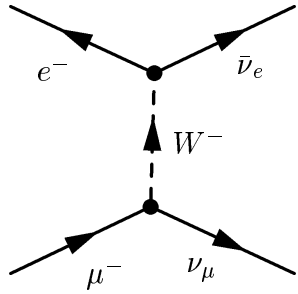}
\end{center}
\caption{The Feynman diagram of the decay  $ \mu^- \to e^- {\bar {\nu}
_e } \nu_\mu $ in the theory with intermediate $W$ boson.}
\label{fig6}
\end{figure}  
Fig.~6 describes the following chain of transitions: the
initial
$\mu^{-}$
emits the final $\nu_{\mu}$ and a virtual $W^{-}$; the vector boson
propagates 
in the virtual state; the virtual 
$W^{-}$-
boson decays into the final $e^{-}$ and
$\bar
\nu_e$. At every vertex the conservation of 4-momenta takes place. 
This ensures the conservation of energy and momentum for the whole 
process.
For a free particle
the square of the 4-momentum is equal to the square of its mass. This is
not
the case for a virtual particle.
For the square of the 4-momentum of the $W$-boson we have $q^2=(p - p')^2$
where $p$ and $p'$ are the 4-momenta of 
$\mu^{-}$
and $\nu_{\mu}$, respectively.
If the mass squared of the $W$ boson $m_{W}^2$
is much larger than  $q^2$ then
the propagator of the $W$-boson (dashed line in
Fig.~6)
gives 
to the matrix element of the process the contribution
proportional to
$\frac{1}{m_{W}^2}$. The diagrams in
Fig.~5 and
Fig.~6
are equivalent if the Fermi constant is connected to the constant
$g$
by the relation

\begin{equation} 
 \frac{G_{\rm{F}}}{\sqrt 2} = \frac{g^2}{8 m^2_W}  
\end{equation}

The universal current $ \times{}$ current theory of the weak
interactions,
as well the theory with the intermediate $W$-boson,
 allowed one
to describe the data of many experiments. Nevertheless both theories
could not be considered as a final theory of
the weak
interactions.
The main reason was that
both theories were not renormalizable quantum field theories.
The probability of transitions calculated in
lowest order perturbation theory were in a good agreement with
experimental data. However, the corrections due to higher orders
of perturbation theory cannot be calculated: they contained divergent
integrals from which it could not be found the finite corrections by the
renormalization of masses and
interaction constants. At that time the only known 
renormalizable
theory, that
allowed to calculate the higher order corrections and that was in
an excellent
agreement with experiment, was
quantum
electrodynamics.

The enormous progress in the understanding of weak interactions 
is
connected with the development of the Glashow-Weinberg-Salam 
renormalizable
theory of weak and electromagnetic interactions, 
the, so called,  Standard Model (SM). We will discuss this
theory later.

\section{Discovery of the $\nu_{\mu}$. Electron and muon lepton numbers}

The mass of the muon is approximately 200 times larger than the electron
mass
($m_{\mu}$=105.66 MeV and $m_{e}$=0.51 MeV). From the very
beginning of the investigation  of muons the possible decay channel
\begin{equation} 
 \mu \to e + \gamma 
\label{MUEG}
\end{equation}
was searched for. No indications in favor of this decay were
found. In the first experiments that were done at the end of the
forties,
for the upper bound of the ratio $R$ of the probability of the decay
$\mu^{+} \to e^{+} + \gamma$ to the probability of the decay 
$\mu^{+} \to e^{+} + \nu_e + \bar \nu_{\mu}$ , which is the main decay
channel of muon, it was found that
$ R < 10^{-2}$. At present the upper bound of $R$ is found to be 
$ R < 1.2\times{10^{-11}}$

If the muon and electron neutrinos are the same particles
the process (\ref{MUEG}) is possible. At the end of fifties 
the probability of the decay
$\mu \to e + \gamma$ was calculated in a nonrenormalizable 
theory with 
$W$-boson
and
the estimated value of the ratio $R$ was larger than 
existed at that time upper
bound ($ R< 10^{-8}$). This was a possible indication that $\nu_e$
and
$\nu_{\mu}$ 
were different particles.
It was necessary, however, to check this in a direct experiment. Such an
experiment
was
proposed by B. Pontecorvo in 1959 and it was done by
L. Lederman, M. Schwarz, J. Steinberger et al in 1962 in Brookhaven 
(USA).

The Brookhaven experiment was the first experiment that has been done with
neutrinos from an accelerator. The beam of pions in this experiment was
produced by
the bombardment of a Be target by 15 GeV protons. Neutrinos were
produced
in the decays
of pions in a decay channel (about 20 m long). 
After the channel
there was an iron shielding  13.5
m thick, in which
charged particles were absorbed. After the shielding there was a neutrino
detector (about 10 tons).  

There are two decay modes of the $\pi^{+}$:

\begin{eqnarray}
\pi^{+} \to \mu^{+} + \nu_\mu
\label{PIMU}
\\
\pi^{+} \to e^{+} + \nu_e
\label{PIE}
\end{eqnarray}

In the Feynman-Gell-Mann theory the decay (\ref{PIE}) is strongly
suppressed.
In fact, let us consider this decay in the rest frame of the pion.
In this frame the $e^{+}$ and the neutrino are moving in opposite
directions.
The helicity of the neutrino is equal to $-1$.
If we neglect the mass of the positron 
the helicity of the positron will be 
equal to $+1$
(the helicity of the positron in this case will be the same 
as the helicity of the antineutrino)
Thus, the projection of the total angular momentum on the direction of the
momentum of the positron will be equal to 1. However,
the spin of
the pion is
equal to zero and the projection of the initial angular momentum on 
any direction is
equal to zero.  
Thus, in the limit $m_e \to 0$ the decay (\ref{PIE}) is forbidden.
For $m_e \neq 0$ the decay (\ref{PIE}) is not forbidden but 
it is 
strongly suppressed with
respect to the decay (\ref{PIMU}). The ratio of the probabilities 
of the decays (\ref{PIE})
and (\ref{PIMU}) is given by

\begin{equation}
R=(\frac{m_e}{m_\mu})^2 \frac{(1-\frac{m_e^2}{m^2_\pi})^2}
{(1-\frac{m^2_\mu}{m^2_\pi})^2}
\simeq{1.2 \cdot 10^{-4}}
\end{equation}

Thus, in decays of pions predominantly muon neutrinos are produced.

In the neutrino detector the processes of the interaction of neutrinos with
nucleons
were observed. If $\nu_{\mu}$ and $\nu_{e}$ are different particles,
muons produced in the process 

\begin{equation}
\nu_\mu + N \to \mu^- +X
\label{NUNM}
\end{equation}
will be observed in the detector ($X$ mean any hadrons). If 
$\nu_{\mu}$ and $\nu_{e}$ are the same particles, the
process

\begin{equation}
\nu_\mu + N \to e^- + X
\label{NUNE}
\end{equation}
is also  possible and in the detector muons {\it and electrons}
will be
observed. Due to $\mu-e$  universality of the weak interaction 
the cross sections of the processes (\ref{NUNM}) and (\ref{NUNE})
will be practically the same and equal
numbers
of muons and
electrons will be observed in the detector.

In the Brookhaven experiment 29 muons were detected. Only 6
electron events were observed. All electron events could be explained as
background events.
Thus, it was proved that the process (\ref{NUNE}) is forbidden,
i.e. {\it
muon
and electron neutrinos are different particles}.

To explain the results of the Brookhaven and other experiments, it 
is necessary to
introduce
two conserved lepton numbers: 
the electron lepton number $L_{e}$ and
the muon lepton number
 $L_{\mu}$. The electron and muon
lepton numbers of different particles are given in the Table 1.

\begin{table}[t]
\begin{center}
\begin{tabular}{|c|c|c|c|} \hline
    & $\nu_e,~e^-~ $ & $ \nu_\mu,~\mu ^- $ & $
\rm {hadrons, \gamma, \ldots} $ \\
\hline
$~~ L_e ~~   $  &  1  &  0  &  0  \\
$~~ L_\mu~~  $  &  0  &  1  &  0  \\
\hline
\end{tabular}
\end{center}
\caption{ \label {Lepton numbers}
Lepton numbers of particles.}
\end{table}

{}From the conservation of the total electron and total muon lepton numbers

\begin{equation}
\sum L_e = {\rm{const}}
\,,
\qquad
\sum L_\mu = {\rm{const}}
\,.
\end{equation}
it follows that the decays

\begin{equation}
\mu^+ \to e^+ \gamma
\,,
\qquad
\mu^+ \to e^+ e^- e^+
\end{equation}
and other similar processes are forbidden.

Let us notice that from 
the modern point of view
the family lepton numbers  $L_{\mu}$ and  $L_{e}$ are violated due to
small neutrino masses and neutrino mixing.
This violation can be revealed in neutrino oscillations that we will
discuss later.

\section {Strange particles in the current $ \times $ current
interaction. The Cabibbo angle}

In the fifties a large family of new particles 
$K^{\pm}$, $K^0$, $\bar K^0$, $\Lambda$, $\Sigma^{\pm,0}$, $\Xi^{-,0}$
was discovered. 
These particles were called strange particles.

Strange particles are produced in nucleon-nucleon and pion-nucleon
collisions
only in pairs. 
For example, the process

\begin{equation}
\pi^- + p \to \Lambda + K^0
\label{PIK}
\end{equation}
in which two strange particles are produced,
was observed . On the other hand, it was shown that the process of
production of
one strange particle

\begin{equation}
n + p \to \Lambda + p
\label{NPP}
\end{equation}
was forbidden.

In order to explain the fact of the production of strange particles in pairs
in nucleon-nucleon and pion-nucleon collisions it was necessary to introduce
a conserved quantum number 
that distinguished  strange particles from nonstrange ones (nucleons,
pions
and others ).
This quantum number was called
{\it strangeness S} . If we assume that  the nucleon and pion have $S=0$,
$K^{0}$ has
$S=1$ and $\Lambda$ has $S=-1$, then the process (\ref{PIK}) is
allowed and the process (\ref{NPP}) is forbidden.

Strange particles are unstable and in their decay the strangeness is not
conserved. The investigation of processes such as 

\begin{eqnarray}
K^+ \to \mu^+ + \nu_\mu, ~~~~ \Lambda \to n + e^- + \bar{\nu}_e ~,
\nonumber \\
\Sigma^- \to n + e^- + \bar{\nu}_e~~~~\Xi^- \to \Lambda + 
e^- + \bar{\nu}_e
\end{eqnarray}
and others allowed to formulate two phenomenological rules that
govern these decays.

I. In decays of strange particles the strangeness is changed by one, i.e.,
 $|\Delta S|=1$.

II. The rule $\Delta Q = \Delta S$ is satisfied 
($\Delta Q = Q_f - Q_i$
and $\Delta S =S_f - S_i$, $ Q_i (S_i)$
and $Q_f (S_f)$ are initial (final) total charge and strangeness of
hadrons).

According to rule I the decay

\begin{equation}
\Xi^- \to \Lambda + e^- + \bar{\nu}_e
\end{equation}
is allowed and the decay
\begin{equation}
\Xi^- \to n + e^- + \bar{\nu}_e
\end{equation}
is forbidden (the strangeness of $\Xi$ is equal to -2).

According to  rule II the decay
\begin{equation}
\Sigma^+ \to n + e^+ + \bar{\nu}_e
\end{equation}
is forbidden (the strangeness of $\Sigma^{\pm}$ is equal
to -1). All these predictions are in perfect agreement with experiments.

In 1964 Gell-Mann and Zweig made the crucial assumption that
the proton,
the neutron,
the pions, the strange particles and all other hadrons are bound states of
{\it
quarks}. Quarks are
particles 
with spin 1/2,
electric charges 2/3 or -1/3 ( in the units of the electric charge of
the proton) and
baryon number equal to 1/3.
Gell-Mann and Zweig introduced
three quarks, constituents of nonstrange and strange 
hadrons: nonstrange quarks
$u$ and $d$ with charges 2/3 and -1/3, respectively and a strange
quark $s$
with charge -1/3 and strangeness -1. In the framework of the quark
model the proton is a bound state of two $u$-quarks and a $d$-quark,
the $\pi^+$ -meson is a
bound state of a $u$-quark and a $\bar d $-antiquark, the $K^+$-meson
is a bound state of
a $u$-quark and a $\bar s$-antiquark, the $\Lambda$- hyperon is a
bound
state of
a $u$-quark, a $d$-quark and a $s$-quark etc. The correctness of the 
quark hypothesis was confirmed
by numerous experiments. Later we will discuss the role of the neutrinos 
in revealing
the quark structure of the nucleon.

If nucleons, pions, strange particles
and other hadrons are not elementary
particles and instead are bound states of quarks it is natural to assume
that
the fundamental weak
interaction is the interaction of leptons, neutrinos  {\it and
quarks}. In this case
the Feynman diagram of the $\beta$-decay of the neutron has
the form
presented in
Fig.~7.
\begin{figure}
\begin{center}
\includegraphics*[height=5cm]{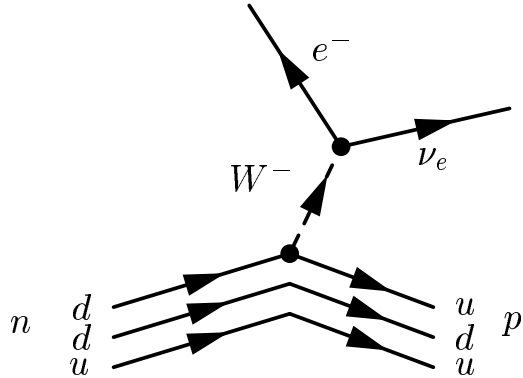}
\end{center}
\caption{The Feynman diagram of the process $n \to p e^- {\bar{\nu}_e}$
in the quark model.}
\label{fig7}
\end{figure}  
Strange particles were included into the current $\times{}$ current
interaction by N. Cabibbo in 1963.
The current $ \bar{p}_L \gamma_{\alpha} n_L $
does not change strangeness and changes charge by one. 
The quark current that
has such properties is
 $ \bar{u}_L \gamma_{\alpha} d_L $.
The quark current that changes charge by one and changes strangeness
is $ \bar{u}_L \gamma_{\alpha} s_L$
This current satisfies rules I and II
{\it automatically}.

It was also known  from  the analysis of experimental data 
that decays of strange particles are suppressed with respect to
the decays of nonstrange particles. To take into account this suppression 
N. Cabibbo
introduced an additional parameter. This parameter is  called
the Cabibbo angle
$\theta_C $. For the quark weak
current he proposed the following expression

\begin{equation}
j^{\rm{C}}_{\alpha}= 2[\cos{\theta_C} 
\bar{u}_L \gamma_{\alpha} d_L + \sin{\theta_C}
\bar{u}_L \gamma_{\alpha} s_L] 
\end{equation}  

It was shown that the weak interaction Hamiltonian with such a current
allows one
to describe experimental data.
{}From the analysis of the data it was found that $ \sin{\theta_C}
\simeq 0.2$.

Let us write down the total weak current in the form
\begin{equation}
j_{\alpha}=2[\bar{\nu}_{eL} \gamma_{\alpha} e_L + 
\bar{\nu}_{\mu L} \gamma_{\alpha} \mu_L +
\bar{u}_L \gamma_{\alpha} d_L']
\label{WCUR}
\end{equation}
where 

\begin{equation}
d'_L = \cos{\theta_C} d_L + \sin{\theta_C} s_L
\end{equation}
is the mixture of the fields of the $ d$ and $s$ quarks.

Notice that there are two
lepton terms
and one quark term in the expression (\ref{WCUR}).
In 1970 it was shown by Glashow, Illiopulos and Maiani that in 
the 
case of the current (\ref{WCUR}) 
the probability of the decays of the type

\begin{equation}
K^+ \to \pi^+ + \nu + \bar{\nu}
\end{equation}
in which $\Delta S =-1$ and $\Delta Q = 0$ is significantly larger than
the
upper bound obtained in experiments. In order to avoid this problem 
Glashow, Illiopulos and Maiani
assumed
that there exists a fourth
quark with  charge 2/3 
and that there is an additional term 
in the weak current
in which the field of the new quark enters. This new quark was called
the charm quark ($c$).
The weak currents took the form

\begin{equation}
j_{\alpha} = 2[\bar{\nu}_{eL} \gamma_{\alpha} e_L +
\bar{\nu}_{\mu L} \gamma_{\alpha} \mu_L +
\bar{u}_L \gamma_{\alpha} d'_L +
 \bar{c}_L \gamma_{\alpha} s'_L]
\end{equation}
where

\begin{eqnarray}
d'_L & = & \cos{\theta_C} d_L + \sin{\theta_C} s_L
\nonumber \\
s'_L & = & -\sin{\theta_C}d_L + \cos{\theta_C} s_L
\end{eqnarray}

The symmetry between leptons and quarks was restored.

In 1976 the first charmed mesons $D^{\pm,0}$, bound states of charmed
and $u$ ($d$) quarks, were
discovered in the
experiments at
$e^{+}-e^{-}$ colliders. Later other charmed mesons and charmed baryons
were also observed.

\section {Glashow-Weinberg-Salam theory of the electroweak
interaction}

The current $\times$ current theory  of the weak interaction and
the theory with heavy charged vector $W^{\pm}$ bosons
to lowest  order  perturbation theory allowed one to describe all
existing 
experimental data.
However, both theories were only effective nonrenormalizable theories: in the
framework of these
theories it was not possible to calculate corrections due to higher orders of
perturbation theory. 

The modern renormalizable theory of the weak interaction (S.L. Glashow
(1961), S. Weinberg (1967) and A. Salam (1968)) appeared as a result of
{\it unification of the 
weak and electromagnetic interactions  into an electroweak
interaction }.This
theory which is called the Standard Model (SM) is one of the greatest
achievements of particle physics in the 20th century. This theory
successfully 
predicted
the existence of families of new hadrons (charmed, bottom and top), 
new interactions
(Neutral Currents), the existence of $W^{\pm}$ and $Z^{0}$ bosons, masses
of these particles
etc. All predictions of the Standard Model are in
perfect
agreement with
existing experimental data including very precise high-energy data that
were
obtained in experiments
at $e^{+}-e^{-}$ colliders at CERN (Geneva) and SLAC (Stanford). 

The Hamiltonian of the electromagnetic interaction has 
the form of the scalar product
of the electromagnetic current and the electromagnetic field

\begin{equation}
{\cal{H}}_I^{\rm{em}} = e \ j^{\rm{em}}_{\alpha} \ A^{\alpha}
\end{equation}

Here

\begin{equation}
j_{\alpha}^{\rm{em}} = 
\sum _{l=e,\mu} (-1) \bar{l} \gamma_{\alpha} l +
\sum _{q=u,d,...} e_q \bar{q} \gamma_{\alpha} q
\end{equation}
is the electromagnetic current of leptons and quarks ($e_{u}=2/3$,
$e_{d}=-1/3, \ldots$ )

The electromagnetic field $A_{\alpha}$ is determined up to the derivative of
an arbitrary function. The observable physical quantities are not
changed
if we
make the following
transformation

\begin{equation}
A_{\alpha}(x) \to  A_{\alpha}(x)- \frac{1}{e} 
\frac{\partial{\Lambda}(x)}{\partial{x^{\alpha}}}
\label{EMPOLE}
\end{equation}
and change correspondingly the unobserved phases of the quark and
lepton fields.
In (\ref{EMPOLE}) $\Lambda (x)$ is an arbitrary function. This
invariance
is called gauge invariance
and the electromagnetic field is an example of a gauge field. A gauge
field is
a vector field and corresponding particles, quanta of the gauge field,
have spin equal to one.

Weak and electromagnetic interactions are unified on the basis of the
generalized Yang-Mills
gauge invariance. The corresponding gauge fields include not only
the electromagnetic field
but also fields of the charged vector particles. 

The SM is based on spontaneously broken SU(2) $\times$ U(1) gauge
symmetry which assumes the existence, in addition to the  massless
photon,
three massive spin 1 particles: two charged and one neutral.
The Hamiltonian of the SM has the following
form

\begin{equation}
{\cal{H}}_I = (\frac{g}{2\sqrt{2}} j_{\alpha} W^{\alpha} + 
h.c.) + \frac{g}{2 \cos{\theta_W}} j^0_{\alpha} Z^{\alpha} +
e j_{\alpha}^{\rm{em}} A^{\alpha}
\label{HSM}
\end{equation}
 
Here

\begin{equation}
j^0_{\alpha} = 2 j^3_{\alpha}- 2 \sin^2{\theta}_W
j_{\alpha}^{\rm{em}}=
\sum_l{\bar{\nu}_{lL}}\gamma_{\alpha} \nu_{lL} + \ldots
\end{equation}
is the so called neutral current and  $ \theta_W $ is a parameter
(Weinberg or weak angle).

The first term of (\ref{HSM}) is the
charged current (CC) interaction, that we have discussed before.
The second term is a new neutral current (NC) interaction.
Third term is the well known electromagnetic interaction.

Thus, the unified theory of the electroweak interaction {\it
predicted
the existence of a new neutral vector boson $Z^{0}$ and a new NC
interaction}.

This new interaction means the existence of new weak interaction
processes.
The first processes were discovered in 1973 at CERN. We will
discuss this discovery
in the next chapter. Charged 
$W^{\pm}$ and neutral $Z^{0}$ bosons were discovered in experiments at
the proton-antiproton collider at CERN in 1983.

\section {The discovery of neutral currents}

Beams of neutrinos (antineutrinos) that can be obtained at
accelerators are mainly the beams
of muon neutrinos (antineutrinos) from decays of pions with a
small (a few \%)
admixture 
of electron neutrinos and antineutrinos from the decays of kaons and
muons.

We will discuss NC processes that were observed in experiments with the
beam of high energy
neutrinos at CERN in the beginning of the eighties.

If the muon neutrino (antineutrino) interacts with a nucleon the
following processes  

\begin{equation}
\nu_\mu (\bar{\nu}_\mu) + N \to \mu^- (\mu^+) + X
\label{NUMX}
\end{equation}
are possible. The diagram of the neutrino process is presented in
Fig.~8.
\begin{figure}
\begin{center}
\includegraphics*[bb = 240 570 135 685, height=5cm]{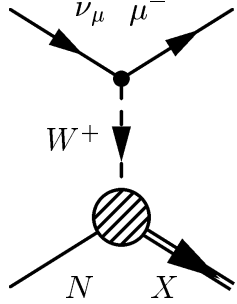}
\end{center}
\caption{The Feynman diagram of the  inclusive process $\nu_\mu + 
N \to \mu^- + X $.}
\label{fig8}
\end{figure}  
This Feynman diagram describes the following steps: due to the CC
interaction (\ref{HSM}) the initial
$\nu_{\mu}$ produces the final
$\mu^{-}$
and 
virtual $W^{+}$ boson;
the virtual $W^{+}$ boson propagates and is absorbed by a quark
inside of the nucleon. As a result 
the initial quark is transferred into the final quark
(the initial
nucleon is transferred into final hadron states).
If only the final muon is observed and the effective mass of the
final
hadrons is much larger than the mass of the nucleon, the process is
called an inclusive
deep inelastic process.

The process (\ref{NUMX}) is a typical weak interaction process:
absorption of a neutrino is {\it accompanied} by the production of
a corresponding
charged
lepton (like in $\beta$-decay of the neutron, the  production of
an electron is
accompanied
by emission of a $\bar\nu_e$).

If there is the NC interaction (\ref{HSM}) the deep inelastic NC processes
  
\begin{equation}
\nu_\mu(\bar{\nu}_\mu) + N \to \nu_\mu(\bar{\nu}_\mu) + X
\end{equation}
with a neutrino (and not a muon) in the final state 
become possible (see diagram
Fig.~9).
\begin{figure}
\begin{center}
\includegraphics*[height=4cm]{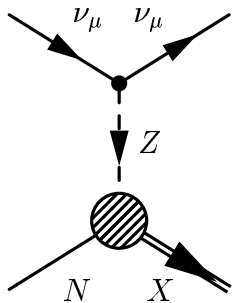}
\end{center}
\caption{The Feynman diagram of the inclusive process  $\nu_\mu +  
N \to \nu_\mu + X $.}
\label{fig9}
\end{figure}
In the Feynman diagram Fig.~9
due to the NC interaction (\ref{HSM}) the
initial
$\nu_{\mu}$ produces the final $\nu_{\mu}$ and a virtual $Z^{0}$
boson.
The virtual $Z^{0}$ boson  propagates and is absorbed by a quark
inside the
nucleon. As a result of this absorption the 
initial nucleon is transferred in a final hadron state.

Such a new weak process was  first observed at CERN in 1973 in the
bubble chamber
"Gargamelle".                                                         
It was found that the ratio of the NC and CC cross
sections is approximately equal to 0.3.
Thus, investigation of neutrino processes
allowed one
to {\it discover new weak processes}. 
The discovery 
of NC processes and their detailed investigation were crucial 
confirmation of
the Glashow-Weinberg-Salam unified theory of the weak
and
electromagnetic interactions.

Another
NC process is the process of elastic scattering of $\nu_{\mu}$ 
($\bar \nu_{\mu})$ on electrons
(see diagram Fig.~10)
\begin{figure}
\begin{center}
\includegraphics*[bb = 240 570 135 685, height=5cm]{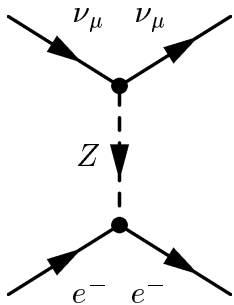}
\end{center}
\caption{The Feynman diagram of the process $\nu_\mu +  
e \to \nu_\mu + e $.}
\label{fig10}
\end{figure}
\begin{equation}
\nu_\mu(\bar{\nu}_\mu) + e \to \nu_\mu(\bar{\nu}_\mu) + e
\label{NUME}
\end{equation}

The cross sections of these processes were measured at high energies by
the CHARM
collaboration at CERN. For the cross sections it was found
\begin{eqnarray}
\sigma_{\nu_{\mu}e} & = & (1.9 \pm 0.4 \pm
0.4) 10^{-42}~\frac{E}{\rm {GeV}}~cm^2
\label{SMUE}
\\
\sigma_{\bar{\nu}_{\mu}e} & = & (1.5 \pm 0.3 \pm
0.4)10^{-42}~\frac{E}{\rm {GeV}}~cm^2
\label{SAMUE}
\end{eqnarray}

{}From these measured cross sections 
the following value of the parameter $\sin^{2}\theta_{W}$
was found
\begin{equation}
\sin^2{\theta_W} = 0.215 \pm 0.032 \pm 0.012.
\end{equation}

This value of the parameter $\sin^2 {\theta_W}$ is in agreement with the
values
obtained from
the measurements of all other NC processes.

Only the NC interaction gives contribution to the
cross sections of the processes (\ref{NUME}).
 The processes
of elastic scattering of the $\nu_e$ and $\bar \nu_e$ on electrons
\begin{equation}
\nu_e(\bar{\nu}_e) + e \to \nu_e(\bar{\nu}_e) + e
\label{NUEE}
\end{equation}
are due to $W$ and  $Z$ exchanges (see diagram Fig.~11).
\begin{figure}
\begin{center}
\includegraphics*[bb = 375 570 135 685, height=5cm]{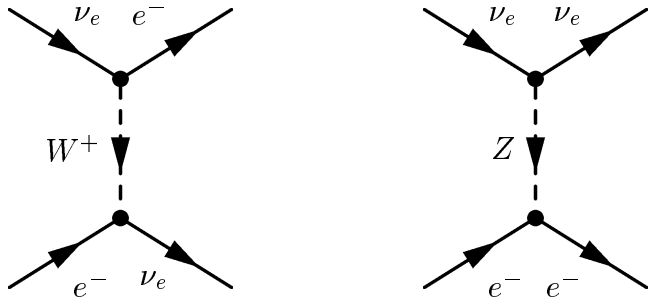}
\end{center}
\caption{The Feynman diagrams of the process  $\nu_e +  
e \to \nu_e + e $.}
\label{fig11}
\end{figure}
Cross sections
of these processes were measured in the experiments at the reactors and at
the
Los Alamos
Meson Factory. Notice that, the CC part of the amplitude of the
 elastic scattering of $\nu_e$ on the electron 
(diagram Fig.~11)  plays a crucial role in the
propagation of neutrinos through matter (see below)

Effects of neutral currents were also measured in the inclusive deep
inelastic
scattering of longitudinally polarized electrons and muons on nucleons
(SLAC and CERN) and 
in atomic transitions. All NC data perfectly confirm the
Standard Model of electroweak interactions. For the parameter
 $\sin^2{\theta_W}$, it was found the value
\begin{equation}
\sin^2{\theta_W} = 0.23155 \pm 0.00019.
\end{equation}

\section{Deep inelastic neutrino-nucleon scattering and quark
structure of the nucleon}

Experiments on the investigation of the deep inelastic CC
neutrino processes

\begin{eqnarray}
\nu_\mu + N & \to & \mu^- + X
\label{DEEPMP}
\\
\bar{\nu}_\mu + N & \to & \mu^+ + X
\label{DEEPMM}
\end{eqnarray}
that have been done at
Fermilab (USA) and CERN in the seventies and eighties
were very important
for establishing the quark structure of the nucleon.
In particle physics these experiments and also the experiments on the deep
inelastic
scattering of electrons (muons) on nuclei played 
the role of the famous
Rutherford experiments
in 
atomic physics. Like the Rutherford
experiments which allowed one to establish the existence of heavy
nuclei
in  atoms,
these experiments allowed one to establish the existence of 
quarks and antiquarks
in nucleons.

Let us first introduce the variables that are usually used to describe
deep inelastic scattering

\begin{equation}
x = \frac{Q^2}{2 p q},~~~
\nonumber \\
y = \frac{p q}{p k},~~~
\nonumber \\
E = \frac{p k}{M}
\end{equation}
where $ q=k-k'$ is the 4-momentum transfer (4-momentum of the
$W$-boson), 
$Q^2 = - q^2$ and $ M $ is the mass of the nucleon.
($ p, k$ and $k'$ are 4-momenta of the initial nucleon, neutrino 
and final muon, respectively). 

{}From conservation of energy and momentum it follows 
that the variable $x$
takes  values in the interval $0\leq x \leq1$. In the lab. system (the
system where the
initial nucleon is at rest)
the variable $y$ becomes

\begin{equation}
y = \frac{E -E'}{E}
\end{equation}
where $E$ and $E'$ are the energies of the initial neutrino and final
muon, respectively. Thus,
$y$ is the relative energy that is transfered to the hadrons. At high
energies
$ 0\leq y\leq1$.
Let us also introduce the variable $\nu =  {pq}/{M}$.
In the region of deep inelastic scattering  $\nu \gg M $ and 
$ Q^{2}\gg M^{2}$. 

Let us consider the processes of interaction of the neutrino with the
$u $ and  $d$ quarks and antiquarks

\begin{eqnarray}
\nu_\mu & + & d \to \mu^- + u
\label{NMU}
\\
\nu_\mu & + & \bar{u} \to \mu^{-} + \bar{d}.
\label{NMD}
\end{eqnarray}

In the deep inelastic region we can neglect the masses of the quarks 
and from conservation of energy and momentum it follows that the virtual
$W$-boson interacts only with those quarks, which have 
momentum $xp$, where $p$ is the nucleon momentum.
The contributions to the differential cross section of the process
$ \nu_\mu + p \to \mu^-+ X $ of the subprocesses (\ref{NMU}) and
(\ref{NMD}) 
are given by the following expression

\begin{equation}
\frac{d^{2}\sigma _{\nu p}}{dxdy} =
2\sigma_0 x[d(x) + (1-y)^2 \bar{u}(x)].
\end{equation}
Here 
\begin{equation}
\sigma_0 = \frac{G^2_F}{\pi}M E \simeq 1.5\cdot 10^{-38}
\frac{E}{\rm {GeV}}~\rm {cm}^2
\label{SECEN}
\end{equation}
is the total cross section of the interaction of the neutrino with
a point-like 
particle with  mass $M$, $d(x)$ and $\bar u(x)$ are
number-densities
of the $d$-quarks and $\bar u $-antiquarks with momentum $xp$ in the
proton.

The dependence of the cross sections on the variable $y$ is determined by
the helicities of the initial particles. Let us consider the process
(\ref{NMD}) in
the center of mass system. In this system the total momentum of the
initial (final) particles
is equal to zero . The helicity of the neutrino is equal to $-1$ and
the helicity of the antiquark $\bar u$ is equal to 1 (we neglect quark
masses).
 Thus, the projection of
the total angular momentum on the direction of the momentum of
the neutrino is
equal
to $2 \times (-{1}/{2})=-1$.  Let us consider the
emission of  a $\mu^-$ in the backward
direction. This
case corresponds to $y=1$ (the energy, that is transferred to
the hadrons, is
maximal). The helicity of the $\mu^-$ is equal to
$-1$ and the projection of the total
angular momentum
on the direction of the momentum of the neutrino is equal to $+1$ in
this case.
Thus, the emission of the $\mu^-$ in the backward direction is
forbidden by 
conservation of total angular momentum. This corresponds to
the $(1-y)^{2}$ dependence of the contribution of the antiquarks to
the cross section of the process (\ref{DEEPMP}).

In the case of the process (\ref{NMU}) the projections of the total
angular
momentum on the direction of the momentum of the neutrinos 
are equal to zero for the initial and final particles.
Thus, emission 
of $\mu^-$ in backward direction is allowed. This corresponds to the
absence of an  
$y$-dependence in the  contribution of quarks to the cross section 
(\ref{SECEN}).

In neutrino experiments the target nuclei are usually 
nuclei with
approximately equal numbers of protons and neutrons.
If we take into account the contribution of only $u$ and $d$ quarks, for
the cross sections, averaged over $p$ and $n$, we obtain the following
expression

\begin{equation}
\frac{d^{2}\sigma_{\nu N}}{dxdy} = \sigma_0 x [q(x) + (1-y)^2 q(x)].
\label{SECNN}
\end{equation}
Here 
\begin{eqnarray}
q(x) & = & u(x) + d(x)
\nonumber \\
\bar{q}(x) & = & \bar{u}(x) + \bar{d}(x)
\end{eqnarray}
Here $u(x)$ is the density of $u$-quarks in the proton ($d$-quarks in the
neutron)
and so on.

For the
averaged
cross section of the process

\begin{equation}
\bar{\nu}_\mu + N \to \mu^+ + X
\label{ANMUX}
\end{equation}
we have 

\begin{equation}
\frac{d^{2}\sigma_{\bar{\nu}}}{dxdy} = \sigma_0 x
[(1-y)^2 q(x) + \bar{q}(x)]
\label{SECANN}
\end{equation}

The expressions (\ref{SECEN}), (\ref{SECNN}) and (\ref{SECANN}) were
obtained in the so-
called
naive quark-parton model in which
interactions between quarks are neglected.
If we take into account the interaction of quarks with gluons in this
case the expressions for the cross sections have the same form, 
but the quark and antiquark distribution functions $q$ and $\bar{q}$
will depend
not only on the variable $x$ but also on $\ln Q^{2}$. 

Expressions (\ref{SECNN}) and (\ref{SECANN}) allows one to describe
existing
experimental
data.
{}From these expressions it is possible to obtain information on the 
distribution of quarks and antiquarks in the nucleon.

 For $y$ - distributions
from (\ref{SECNN}) and (\ref{SECANN}) we have
\begin{eqnarray}
\frac{d\sigma_{\nu N}}{dy} & = & 
\sigma_0 [Q + (1-y)^2 \bar{Q}]
\nonumber \\
\frac{d\sigma _{\bar{\nu}N}}{dy} & = & 
\sigma _0 [(1-y)^2 Q + \bar{Q}]
\label{SEQAQ}
\end{eqnarray}
where

\begin{equation}
Q = \int_{0}^{1} x q(x) dx,~~~~~\bar{Q} = \int_{0}^{1} x\bar{q}(x) dx
\end{equation}
are the fractions of the momentum of the nucleon carried by quarks
and antiquarks, respectively (in the system $q^{0} =0$, in which the
momentum of the nucleon is much larger than its mass)

{}From the relations (\ref{SEQAQ}) it follows that at $y=0$ the cross
sections of the
processes (\ref{DEEPMP}) and (\ref{DEEPMM}) must be equal. 
This is  confirmed by the data of the neutrino experiments.
{}From the data of the CDHS
experiment at CERN with neutrino energies in the range 
$ 30 < E < 200 \rm {GeV} $ 
it was found that

\begin{equation}
\left(\frac{d\sigma_{\bar{\nu}N}}{dy}\right)_{y=0}
\
\Bigl/
\
\left(\frac{d\sigma_{\nu N}}{dy}\right)_{y=p}
\
= 1.01 \pm 0.07.
\end{equation}

If the contribution of antiquarks into the cross sections are
much less than the contribution of quarks, we must expect
weak dependence of the cross section $ \frac{d\sigma_{\nu N}}{dy} $ 
on the $y$ and $(1-y)^2$-dependence of the cross section 
$\frac{d\sigma_{\bar{\nu}N}}{dy} $. This behavior was observed in
experiments.
{}From the  analysis of the
CDHS data it follows

\begin{equation}
\frac{\bar{Q}}{Q + \bar{Q}} = 0.15 \pm 0.01
\end{equation}

Thus, the contribution of antiquarks to the nucleon momentum is
about
15 \% of total contribution of the quarks and antiquarks.

For the fraction of the nucleon momentum that is carried by quarks
and antiquarks it
was found that
\begin{equation}
Q + \bar{Q} = 0.492 \pm 0.006 \pm 0.019
\end{equation}

Thus, neutrino experiments proved that not all the nucleon 
momentum 
is carried by the quarks and antiquarks. The other part of
the nucleon momentum is carried by the gluons, vector particles that
interact
with
quarks.

Finally, from the quark-parton model it follows that the total
neutrino
and antineutrino cross sections  depend linearly on neutrino
energy
$ E$. 

\begin{eqnarray}
\sigma_{\nu N} & = & \frac{G^2}{\pi} M ( Q + \frac{1}{3}\bar{Q}) E
\nonumber \\
\sigma_{\bar{\nu}N} & = & \frac{G^2}{\pi} M (\frac{1}{3}Q + \bar{Q}) E  
\end{eqnarray}

The data of the experiments perfectly confirm this prediction of the
theory:

\begin{eqnarray}
\sigma_{\nu N} & = & (0.686 \pm 0.019) \times 10^{-38}
\frac{E}{\rm{GeV}} \rm{cm}^2
\nonumber \\
\sigma_{\bar{\nu}N} & = & (0.339 \pm 0.010) \times 10^{-38}
\frac{E}{\rm{GeV}} \rm{cm}^2
\end{eqnarray}

Thus, the investigation of the high energy neutrino processes 
allowed one to establish the quark structure of the nucleon
and to obtain important information on the distribution functions of 
quarks and antiquarks in the nucleon.

\section{Neutrino masses. Introduction}

{}From all existing data it follows that
the interaction of neutrinos with
matter is given by the Standard Model,   
However, neutrino masses, neutrino magnetic moments
and other fundamental neutrino properties are basically unknown.
We now come to the problem
of {\it the neutrino masses and neutrino mixing}.

The brief history  of neutrino masses is the
following. Pauli introduced
the neutrino as a particle with a mass (as a constituent of
nuclei). He 
thought  that the 
mass of the neutrino is less than the electron mass.
Fermi and Perrin proposed the first method of measuring the neutrino
mass based on the measurement
of the shape of the high
energy part of the $\beta$-decay spectrum. This
part of the spectrum is due to the emission of a neutrino with
small energy and effects of the neutrino mass in that part of the
spectrum is
the most pronounced. 
In experiments on the determination of the neutrino mass by the 
this
method,
the decay of tritium

\begin{equation}
{}^3{\rm{H}} \to {}^3{\rm{He}} + e^- + \bar{\nu}_e
\end{equation}
is usually investigated.

In the first experiments that were done in the forties  no
effects of a
neutrino mass were seen. From these experiments it was found that the
upper
bound of the 
neutrino mass is much less than the electron mass: 
\begin{equation}
m_\nu < 500~ \rm{eV}.
\end{equation}

With the improvement of experimental technique 
this upper bound became much smaller and
at the time, when the parity violation in $\beta$-decay was
discovered, the upper bound of the neutrino mass was about 100 eV. 

The theory of the two-component neutrino (Landau, Lee and Young and Salam)
was
based on the {\it assumption}
that the
neutrino mass is equal to zero. After the success of this theory during
many years there was a general
belief that all neutrinos are massless particles. The 
Glashow-Weinberg-Salam theory was also based on this assumption.

In 1957-58 B. Pontecorvo considered the possibility of
{\it a small but nonzero neutrino masses}.
The only known massless particle is the photon.
There is a symmetry reason for the photon to be massless-the gauge
invariance of 
quantum electrodynamics. B. Pontecorvo put attention that there is no
such a principle in the case of the neutrino.
He showed that, if states 
of neutrinos produced in weak decays
are  superpositions of the states of 
neutrinos with small masses,
{\it neutrino oscillations} will take place in the beams of the neutrinos
in vacuum,  
similar to well known $K^{0} \to \bar K^{0}$ oscillations.
B. Pontecorvo showed that the search for neutrino oscillations is a
very
sensitive method of the measurement of small neutrino
masses.

In 1962  at the time of the Brookhaven experiment Maki, Nakagawa and
Sakata 
proposed some model in which the
nucleon was considered as a bound state of some vector particle and
massive neutrinos. They assumed  that the fields of $\nu_e$ and
$\nu_{\mu}$ are
linear orthogonal combinations of the fields of the massive neutrinos
and pointed out that in such a case transition of muon neutrinos
into 
electron
neutrinos becomes possible.

In the seventies in Dubna (Russia) and other places in the framework
of the
SM 
the neutrino masses and mixing were considered as a phenomena 
analogous to the Cabibbo-GIM quark mixing. The neutrino oscillations
between two
types of
neutrinos were discussed and the different experiments 
on the search for neutrino oscillations were proposed.

At that time the majority of physicists still believed that neutrinos
are massless particles.
The opinion about the neutrino masses drastically changed in the
end of the seventies  with the appearance of 
models beyond the Standard Model such as models of
Grand Unification. These models are based
on 
large symmetry groups and fields of
neutrinos enter into the same multiplets of the groups 
as the fields of leptons and quarks.
A mechanism
of the
generation of the masses of quarks and leptons generally  provides also
masses to the neutrinos.
The neutrino masses and mixing started to be considered as phenomena
connected with physics beyond the Standard Model. 

In the eighties  special experiments on the search for neutrino
oscillation started. The problem of the neutrino masses and neutrino
oscillations became the most challenging and important problems of 
neutrino
physics.

\section{Discovery of the $\tau$-lepton, $b$ and $t$-quarks. The
number of flavor
neutrinos}

Up to now we have considered four leptons: the two charged leptons $e$
and
$\mu$ and the two neutrinos $\nu_e$ and  $\nu_{\mu}$.
In 1975 the third heavy charged lepton $\tau$ with a mass of about 1.8 GeV
was
discovered  by M. Perl et al. at the $ e^+ -e^- $ collider at Stanford
(USA).

In the framework of the SM
this was a discovery of
the third family of leptons and quarks. It meant that a new type of
neutrino $\nu_{\tau}$ and two new quarks with charges $2/3$ and $-1/3$
must
exist. These quarks were called the top and bottom. The real triumph
of the
Standard model was the discovery  of the bottom particles in
the eighties and
top quark in the nineties.

After these discoveries the charged current of leptons and quarks
took the form

\begin{equation}
j_\alpha^{CC} = 2 (\sum_{l=e,\mu,\tau} \bar{\nu}_{lL}
\gamma_{\alpha} l_L
+\overline{u}_L \gamma_\alpha d'_L
+ \overline{c}_L \gamma_\alpha s'_L
+\overline{t}_L \gamma_\alpha b'_L)  
\end{equation}
where
\begin{equation}
d'_L = \sum_{q=d,s,b} V_{uq} q_L,
\qquad s'_L = \sum_{q=d,s,b} V_{cq} q_L,
\qquad b'_L = \sum_{q=d,s,b} V_{tq} q_L
\end{equation}

Here $V$ is the unitary matrix that is called
the Cabibbo-Kobayashi-Maskawa 
matrix. 
The elements of this matrix are well- 
known from the data of numerous experiments.

How many families of quarks and leptons exist in Nature? 
The investigation of neutrino processes allowed to answer this fundamental
question.
As we 
have seen, the number of families is equal to the number of  
neutrino types (neutrino flavors). This number was
measured in 
experiments at the $e^{+}-e^{-} $
colliders at SLC (Stanford) and LEP (CERN). From the data of these
experiments 
the probability (width) of the decay
\begin{equation}
Z \to \nu_l + \bar{\nu}_l~~~~~~l=e,\mu,\tau, \ldots
\label{ZNAN}
\end{equation}

was determined. The width of the decay (\ref{ZNAN}) is proportional
to the number of neutrino flavors $ n_\nu $. From the data of the
recent
LEP experiments it was found that
\begin{equation}
n_\nu =2.994 \pm 0.012.
\end{equation}

Thus, only three flavor neutrinos $\nu_{e}$, 
$\nu_{\mu}$, $\nu_{\tau}$
 and, consequently, three  
families of quarks and leptons exist
in Nature.

\section{Neutrino mixing}

If the neutrinos are massless, the Standard 
weak interaction conserve three lepton numbers $L_e$, $L_{\mu}$ and
$L_{\tau}$:

\begin{equation}
\sum L_e =const,~~~ \sum L_\mu =const,~~~ \sum L_\tau =const
\label{cos}
\end{equation}

The values of the lepton numbers of the charged leptons and the neutrinos
are given in Table \ref{lepnumb}.

\begin{table}[t]
\begin{center}
\begin{tabular}{|c|c|c|c|c|} \hline
    & $\nu_e,~e^-~ $ & $ \nu_\mu,~\mu ^- $ & $  \nu_\tau,~\tau^-$ & $
\rm {hadrons, \gamma, \ldots} $ \\
\hline
$ L_e    $  &  1  &  0  &  0  &  0  \\
$ L_\mu  $  &  0  &  1  &  0  &  0  \\
$ L_\tau $  &  0  &  0  &  1  &  0  \\ \hline
\end{tabular}
\end{center}
\caption {\label{lepnumb}
Lepton numbers of neutrinos and charged leptons.}
\end{table}


We will now assume that the
neutrinos are massive and the lepton
numbers
are
violated by {\it a neutrino mass term}.
In this case fields of neutrinos $\nu_{eL}$, $\nu_{\mu L}$ and
$\nu_{\tau L}$ in the Lagrangian of the weak interaction
will be linear combinations of the fields of neutrinos with definite
masses

\begin{equation}
\nu_{{l}L} = \sum_{i=1,2,3} U_{{l}i} \, \nu_{iL} \,~~~~(l=e,\mu,\tau)
\label{NFIELD}
\end{equation}

Here $U$ is unitary matrix $ (U U^{+}=1) $ and $\nu_{i}$ are
the fields of
neutrinos with masses
$m_{i}$.

Before we will come to the discussion of the consequences of neutrino
mixing (\ref{NFIELD})
let us notice that there are two types of particles with spin
1/2:
Dirac particles and Majorana particles. 

{\it Dirac particles} possess some
conserved charges.
Every Dirac particle has an antiparticle, 
the particle with the same mass and spin but opposite charge.
The electron and the proton are examples of Dirac particles.
Corresponding antiparticles are the positron and the antiproton.

Other possible particles with spin 1/2 are 
{\it Majorana particles }.
All charges of the Majorana particles are equal to zero. Thus, 
a Majorana particle and a Majorana antiparticle
are identical. Up to now Majorana particles were not observed.
The massive neutrinos and  neutralinos, particles predicted by
models of supersymmetry, are possible candidates. Neutral bosons
such as the photon, $\pi^{0}$ and other are 
well known neutral
particles with integral spin that are
identical to
their antiparticles.

There are two possibilities of the violation  
of the lepton number conservation low:

I. The lepton numbers $ L_e$, $L_{\mu}$ and $L_{\tau}$ are violated
separately but the total lepton number
$L=L_e + L_{\mu} + L_{\tau}$ is conserved

\begin{equation}
\sum L = const
\label{Diracsum}
\end{equation} 

In this case the neutrinos $\nu_i$ 
are Dirac particles that possess lepton number $L=1$. The lepton
number 
of the antineutrinos $\bar \nu_{i}$ is equal to $-1$.
The Dirac neutrino masses and neutrino mixing   
can be generated in 
the framework of the SM by the same
mechanism that is responsible for the generation of the masses
and mixing of quarks.

II. There are no conserved lepton numbers.

In this case the
massive neutrinos $\nu_i$ are Majorana particles.
The Majorana neutrino masses and mixing can be generated only in the
framework
of the models beyond the SM.

If massive neutrinos are Majorana particles 
there exist a plausible mechanism of the generation 
of neutrino masses that
connect
the smallness of neutrino masses with the violation of lepton
numbers at a mass scale $M$ that is much larger than the masses of
leptons and
quarks.
This is the so-called  see-saw mechanism.  
The masses of neutrinos are given in the see-saw case by the relation

\begin{equation}
m_i  \simeq \frac {(m^i_{\rm{f}})^2} {M} \ll  m^i_{\rm{f}}
\qquad (i=1,2,3) \,.
\end{equation}
where $m^i_{\rm{f}} $ is the mass of the lepton or quark in the
$i$th family ($ i=1,2,3$).
Let us notice that in the see-saw case the neutrino masses satisfy
the hierarchy 
relation

\begin{equation}
m_1 \ll m_2 \ll m_3
\end{equation}
that follows from the hierarchy of masses of the leptons (quarks) of
the different 
families. 

\section {Neutrino oscillations}

If there is the neutrino mixing

\begin{equation}
\nu_{{l}L} = \sum_{i=1}^{3} U_{{l}i} \, \nu_{iL} \,
\label{mixing1}
\end{equation}
where $\nu_{i}$ is the field of a neutrino (Dirac or Majorana) with  
mass $m_{i}$, for 
the state vector of flavor neutrinos $\nu_{e}$, 
$\nu_{\mu}$ and $\nu_{\tau}$ (neutrinos that are produced in weak decays
and take part in CC or NC neutrino reactions)
with momentum
$\vec{p} $
we have 

\begin{equation}
|\nu_l \rangle = \sum_{i=1}^3 U_{li}^* \, | i\rangle \,.
\label{FDM}
\end{equation}
where  $|i\rangle$ is the state vector of neutrino 
with mass $ m_i $ and energy
$ E_i = \sqrt{m^2_i + \vec{p}^2} \simeq{p + 
\frac{m_i^2}{2 p}}, (m_i^2 \ll p^2)$.
Thus, in the case of neutrino mixing the state of flavor neutrino
is
{\it a superposition of the states of neutrinos with different
masses}.

The relation (\ref{FDM}) is based on the assumption that the mass
differences
of neutrinos are so small that they cannot be revealed in the 
experiments on the investigation of
processes of
neutrino production and detection. The neutrino
mass
differences can be revealed in the neutrino oscillation experiments,
special
experiments 
with a large  {\it macroscopic
distance} between the neutrino source and the neutrino detector.

Let us assume that at $t=0$ the neutrino $\nu_{l}$
was produced ( $l=e, \mu, \tau $ ). At time $t$ 
we have for neutrino state

\begin{equation}
|\nu_l\rangle_t =
\sum_{i=1}^{3} U_{li}^* \, e^{ - i E_i t } \, |i\rangle \,.
\label{state-t}
\end{equation}

 The state
$|\nu_{l}\rangle_{t}$
is the superposition of the states of {\it all} neutrinos $\nu_{e}$,
$\nu_{\mu}$
and $\nu_{\tau}$

\begin{equation}
|\nu_l\rangle_t =
\sum_{l'=e,\mu,\tau} |\nu_{l'}\rangle
{\cal A}(\nu_l \to \nu_{l'}\rangle\,,
\label{state-t2}
\end{equation}
where
\begin{equation}
 {\cal A}(\nu_l \to  \nu_{l'}) = \sum_{i=1}^3
 U_{l' i}e^{-iE_it}U_{l i}^* 
\label{state-t30}
\end{equation}
is the amplitude of the transition  $\nu_{l} \to \nu_{l'}$
for time $t$. For the transition probability we have

\begin{equation}
P_{\nu_l \to \nu_{l'}} = 
\left| \delta_{l'l} +
 \sum_{i} U_{l' i} \, 
 \left( e^{- i \,
 \Delta{m}^2_{i1}\, \frac {L}{2 p}} - 1 \right) \, U_{li}^* \right|^2.
\end{equation}

Here $ L\simeq t$ is the distance between the neutrino source and
detector,
and $ \Delta m^{2}_{i1} = m^{2}_{i}-m^{2}_{1}$ (we have assumed 
that $m_{1} < m_{2}
< m_{3}) $.

Thus, the transition probabilities depend on the ratio $ L/p $.
If for all neutrino mass squared differences 

\begin{equation}
\Delta m^2_{i1}~\frac{L}{2p} \ll 1,
\end{equation}
in this case 
$ P ( \nu_{l}\to \nu_{l'} ) = \delta _{l' l}$ (no transitions between
different flavor neutrinos). 

In the simplest case of transitions between two types of neutrinos
the mixing matrix has the form

\begin{equation}
U = \left(\begin{array}{rr} \displaystyle
\cos\theta \null & \null \displaystyle \sin\theta
\\ \displaystyle - \sin\theta \null & \null \displaystyle
\cos\theta \end{array}\right)
\label{two07}   
\end{equation}
where $ \theta $
is the mixing angle (if $ \theta =0 $
there is no mixing). For the transition probability we have in this case

\begin{equation}
{\rm P}(\nu_l \to \nu_{l'}) =
{\rm P}(\nu_{l'} \to \nu_l)
= \frac {1} {2} \sin^2 {2\theta} (1 - \cos \frac{\Delta m^2 L}
{2p})
\label{PNAN}
\end{equation}
where $ l'\neq l $ and $ l, l' $ take the values ($ \mu, \tau $) or
($ \mu, e $) or ( $ e, \tau $) and $\Delta m^2 = m^2_2 - m^2_1 $. For
the survival probability
we have

\begin{equation}
{\rm P}(\nu_l \to \nu_l) = {\rm P}(\nu_{l'} \to
\nu_{l'}) =
 1 - \frac {1}{2} \sin^2 2\theta (1 - \cos \frac {\Delta m^2 L} {2p})
\label{wanda}
\end{equation}

The expression (\ref{PNAN}) and (\ref{wanda}) can be rewritten in the
form

\begin{eqnarray}
{\rm P}(\nu_l \to \nu_{l'}) = \frac {1}{2} \sin^2
2\theta
\left(1 - \cos\,  2.53 \Delta m^2\frac {L} {E}\right)
\\
{\rm P}(\nu_l \to \nu_l) =    
 1 - \frac {1}{2} \sin^2 2\theta \left(1 - \cos\,  2.53 \Delta
m^2\frac {L} {E}\right)
\label{PSIN}
\end{eqnarray}
where $L$ is the distance in  m ,  $E\simeq p $ 
is the neutrino energy in MeV and 
 $ \Delta m^{2}$ is neutrino mass squared difference in $\rm {eV}^2$.
Thus, the transition probability is the periodical function of the
parameter
$L/E$ .

 Let us consider the 
$ \nu_{\mu} \to \nu_{\tau} $ transitions and assume that
$ \sin^{2} 2\theta =1 $ (maximal mixing).
The
$ \nu_{\mu} \to \nu_{\mu} $
survival probability is equal to one at the points
$ (\frac{L}{E})_1 = \frac{\pi}{2.53\Delta m^2} 2n $ 
$ (n=0,1,2 \ldots)$, and we will find at these points only $ \nu_\mu
$. 
At the values
$ (\frac{L}{E})_2 = \frac{\pi}{2.53 \Delta m^2}(2n + 1) $ 
the survival probability is
equal to
 zero, and only the $ \nu_\tau $ will be found at these points.  
 At all other values of $ L/E $
 we will find 
$\nu_{\mu} $ {\it and} $\nu_{\tau} $. 
 It is obvious that the
 sum of probabilities to find $\nu_{\mu} $ and $\nu_{\tau} $ 
is equal to one.
 
 The phenomena we have described is called {\it neutrino oscillations }.
 In order to observe neutrino oscillations it is necessary that
the mixing
 angle is large enough and the parameter $\Delta m^{2}$ satisfies
 the following condition

\begin{equation}
\Delta m^2 \geq \frac{E}{L}
\end{equation}

The sensitivities to the parameter $\Delta m^{2}$ 
of neutrino experiments at different facilities are quite different
and cover a very broad range of values of $\Delta m^2 $.
The experiments with accelerator neutrinos have
sensitivities to the parameter $\Delta m^{2}$ in the range 
$10-10 ^{-3}~\rm{eV}^{2}$,
the experiments with the reactor neutrinos in the range 
$10^{-2}-10 ^{-3}~\rm{eV}^{2}$,
the experiments with the atmospheric neutrinos in the range
$10^{-1}-10 ^{-4}~\rm {e}V^{2}$ and finally experiments with the solar
neutrinos have sensitivity to the parameter $\Delta m^2$ down to
$10^{-10}-10 ^{-11}~ eV^{2}$

It is convenient to 
introduce the neutrino oscillation length
\begin{equation}
L_0 =  4 \pi \frac {E} {\Delta m^{2}}
\end{equation}  

For the the transition probability we have
\begin{equation}
{\rm P}(\nu_l \to \nu_{l'}) = \frac {1}{2} \sin^2
2\theta
\left(1 - \cos 2\pi \frac {L}{L_0}\right)~~~~~~(l \neq l').
\end{equation}

The expression for the oscillation length can be written in the form

\begin{equation}
L_0 =  2.47  \frac {E ({\rm{MeV}})}{\Delta m^2 ({\rm{eV}}^2)}
\,{\rm{ m}}
\end{equation}  

Neutrino oscillations can not be observed if the oscillation length
is much larger than the distance $L$ between the neutrino source and
the neutrino
detector. In order to observe neutrino oscillations,
oscillation
length must be smaller or of the order of magnitude of $L$.

Let us notice that for the comparison of neutrino oscillation theory with
experimental data
it
is necessary to average the corresponding theoretical expression
for transition probabilities over the neutrino
energy spectrum, the region where
neutrinos were produced and so on. As a result of such averaging, 
the cosine term in the expressions (\ref{PSIN}) usually  disappears.

\section {Experiments on the search for neutrino oscillations}

There are at present data of numerous experiments on the search for
neutrino oscillations. The important indications in favor of
the neutrino masses and
mixing were found in the solar neutrino experiments. The 
compelling
evidence 
in favor
of neutrino oscillations was obtained recently in 
the Super-Kamiokande  atmospheric neutrino 
experiment. Some indications in favor of 
$\nu_{\mu}  \to \nu_{e} $ oscillations 
were found also in the  Los Alamos
accelerator neutrino experiment. In many experiments with
accelerator and
reactor neutrinos no indications in favor of neutrino oscillations were
found. We will first discuss the solar neutrino experiments.

\subsection{The solar neutrino experiments}

The energy of the sun is generated in the reactions of
the thermonuclear
$pp$ and
$CNO$ cycles. From the thermodynamical point of view the energy of the
sun is
produced in the transition of four protons and two electrons into
$^{4} He$
and two neutrinos

\begin{equation}
4 \, p + 2 \, e^- \to\,{^4}{\rm{He}} + 2 \, \nu_e \,,
\label{TRANSIT}
\end{equation}

Thus, the generation of energy of the sun is {\it accompanied by
the
emission of electron neutrinos}

The main sources of solar neutrinos are the reactions that are
listed in 
Table~\ref{sources}. In this table the maximal neutrino energies 
and neutrino fluxes,
predicted by the Standard Solar Model (SSM), are also given.

\begin{table}[t]
\begin{center}
\renewcommand{\arraystretch}{1.45}
\begin{tabular}{|c|c|c|}
\hline
Reaction  
&
\begin{tabular}{c}
Maximal energy 
\\
(MeV)
\end{tabular}
& 
\begin{tabular}{c} 
Standard Solar Model flux
\\
$({\rm{cm}}^{-2}{\rm{s}}^{-1})$ \\
\end{tabular}
\\
\hline
$p\,p \to d\,e^+\,\nu_e$ & $\leq 0.42 $ &
$6.0 \times 10^{10}$ \\
\hline
$e^{-}\,{^7}{\rm{Be}} \to \nu_e\,{^7}{\rm{Li}}$  
& $0.86 $ & $4.9 \times 10^9$ \\
\hline
${^8}{\rm{B}} \to {^8}{\rm{Be}}\,e^+ \,\nu_e$
& $\leq 15 $ & $5.0\times 10^6$ \\
\hline
\end{tabular}   
\end{center}
\caption{ \label{sources} 
Main sources of solar $\nu_e'$ s.}
\end{table}

As it seen from Table~\ref{sources},
solar neutrinos are mainly
low energy 
$pp$ neutrinos. According to SSM the flux of the medium energy
monochromatic
$^{7} Be$
neutrinos is about 10 \% of the total flux. The flux of the high
energy 
$^{8} B$ neutrinos is only about $10^{-2}$ \% of the total flux.
The  $^{8} B$ neutrinos
give, however, the main contribution to the event rates of experiments
with high
energy threshold.

The results of the five underground 
solar neutrino experiments are available at present.
In the pioneering radiochemical experiment by R. Davis et al (Homestake 
mine, USA),
a tank 
filled with 615 tons of 
$C_{2}Cl_{4}$ liquid is used as a target. Solar neutrinos are
detected
in this experiment by a radiochemical method, proposed by
B.Pontecorvo in 1946, through the observation of
the reaction  
\begin{equation}
\nu_{e} + {^{37}\rm{Cl}}\to e^{-} + {^{37}\rm{Ar}} 
\label{NUCLAR}
\end{equation}

The radioactive atoms of $^{37} Ar$ are extracted from the tank by purging it
with $^{4} He$
gas. The atoms of $^{37} Ar$ are placed in a low background proportional 
counter
in which the process

\begin{equation}
e^{-} + {^{37}\rm{Ar}}\to \nu_{e} + {^{37}\rm{Cl}} 
\end{equation}
is observed by the detection 
the Auger electrons (electrons of conversion).

After 2 months of exposition about 16  atoms  of the $ ^{37}$ Ar are
extracted
from
the volume that contains $ 2.2 \times 10^{30} $ atoms  of $^{37}Cl$!

The solar neutrinos have been observed in the Davis experiment for about
30 years. For the
observed event rate $ Q_{Cl}$, averaged over 108 runs, the following
value was obtained 

\begin{equation}
Q_{\rm{Cl}} = 
2.56 \pm 0.16 \pm 0.16 \, ~~\rm{SNU} 
\end{equation}
where 1 $\rm {SNU} = 10^{-36} $ events/atom s.
The observed event rate is about three times less than the rate
predicted by the SSM
\begin{equation}
(Q_{Cl})_{SSM} = 7.7 \pm 1.2 \, ~~\rm{SNU} 
\end{equation}

The minimal neutrino energy at which the process (\ref{NUCLAR}) become
possible
(the threshold of the process) is equal to $ E_{th} = 0.81 $ MeV.
Thus, the low energy $pp$ neutrinos are not detected in the Davis
experiment.
The most important contribution to the event rate comes from the high
energy  $^{8} B$
neutrinos. About 15\% of the events are due to $^{7}Be$ neutrinos.

In the radiochemical GALLEX (Italy) and SAGE (Russia)
experiments the solar
$\nu_{e}$'s  are detected
through the observation of the reaction

\begin{equation}
\nu_{e} + {^{71}\rm{Ga}}\to e^{-} + {^{71}\rm{Ge}}
\label{NUGAG}
\end{equation}

In the GALLEX experiment the target is a tank with 30.3 tons 
of the $^{71}\rm {Ga}$ in the 
gallium-chloride solution. In the SAGE experiment 
a metallic $^{71}\rm{Ga}$ target is used (57 tons of $^{71}\rm{Ga}$ ).

The threshold of the process (\ref{NUGAG}) is $ E_{th} = 0.23 \rm{MeV} $.
Thus, neutrinos from all  
solar neutrino reactions are detected in these experiments (according to the
SSM the contributions of the $pp$ , $^{7}\rm{Be}$ and $^{8}\rm{B}$
neutrinos
to 
the event
rate in the gallium experiments are about 54 \%, 27\% and 10\%,
respectively).
The event rates obtained in the GALLEX and SAGE experiments are
equal

\begin{eqnarray}
Q_{\rm{Ga}} & = & 77.5 \pm 6.2 {}^{+4.3}_{-4.7}\,
~~\rm{SNU}~~~(\rm{GALLEX}) 
\nonumber \\
Q_{\rm{Ga}} & = & 66.6 \pm {}^{+6.8~+3.8}_{-7.1~-4.0} 
~~\rm{SNU}~~~(\rm{SAGE}) 
\end{eqnarray}

The predicted rate is about two times larger than the observed rates

\begin{equation}
(Q_{\rm{Ga}})_{\rm{SSM}} = 129 \pm 8 \,
~~\rm{SNU}
\end{equation}

In the Kamiokande and Super-Kamiokande  
experiments (Japan) the solar neutrinos are detected through the
observation of the process
\begin{equation}
\nu + e \to \nu + e
\end{equation}

In the Super-Kamiokande experiment a large 50 ktons water-Cerenkov 
detector is used.
The inner surface of the detector is covered with 11146 large photomultipliers
in which the Cerenkov light from the  recoil electrons is
detected. About 14
neutrino
events per day are observed by the Super-Kamiokande 
experiment
(in the previous  Kamiokande experiment
one neutrino event per day was detected).
At high energies
the direction of the momentum of the recoil electrons 
is practically the same as the direction of the momentum of the neutrinos.
Thus, the measurement of the direction of the  
momenta of the electrons allows one to detect events induced by
neutrinos coming from the sun. 
 The recoil electron energy threshold 
is rather large (7 MeV in the 
Kamiokande experiment and 5.5 MeV in the
Super-Kamiokande
experiment). 
Thus,  only the $^{8} B$ 
neutrinos are detected in these experiments.
{}From the results of
the Kamiokande and 
Super-Kamiokande
experiments the following values of the solar neutrinos fluxes were
obtained, 
respectively

\begin{eqnarray}
\Phi & = & (2.80 \pm 0.19 \pm 0.33)~10^6
{\rm{cm}}^{-2}{\rm{s}}^{-1}
\nonumber \\
\Phi & = & (2.44 \pm
0.05{}^{+0.09}_{-0.07})~10^6{\rm{cm}}^{-2}{\rm{s}}^{-1}
\end{eqnarray}

The measured fluxes are about 1/2 of the predicted one by the SSM

\begin{equation}
\Phi_{\rm{SSM}} = (5.15 {}^{+1.00}_{-0.72})~10^6
{\rm{cm}}^{-2}{\rm{s}}^{-1}
\end{equation}

Thus, from the results of all solar neutrino experiments 
it follows that the fluxes of 
the solar $\nu_{e}$'s
on the earth 
in different ranges of energies
are significantly smaller than the predicted fluxes.
This deficit constitutes {\it the solar neutrino problem}.

Neutrino oscillations is
the most plausible explanation of the solar neutrino 
problem.
If neutrinos are massive and mixed, the solar $\nu_{e}$'s on the way 
to the earth can be 
transfered into other neutrinos ($\nu_{\mu}$ and/or $\nu_{\tau}$).
However, in the chlorine and gallium experiments
only $\nu_{e}$'s can be detected. 
The muon and/or tau neutrinos give some contribution to the event
rates of the 
Kamiokande and the
Super-Kamiokande experiments. However,
cross section of  $\nu_{\mu} (\nu_{\tau}) -e $ scattering is
about
1/6 of the cross section of  $ \nu_{e}-e$ scattering, and
therefore, 
the main contribution
to the event rate of these experiments also comes from
$\nu_{e}$'s.
Thus, if there are neutrino oscillations,    
the event rates detected in the solar neutrino experiments
will be less
than the expected ones.

Solar neutrinos, produced in the central zone of the sun, on their way
to the earth pass through a large amount of matter of the sun. At some
values of
the mixing parameters effects of the coherent interactions of neutrinos
with 
matter can enhance significantly
the probability of the transition of
solar  $\nu_{e}$'s into other states. 

The refraction index of the neutrinos in matter depends on the
amplitude
of elastic scattering of neutrinos in the
forward direction.
Both the CC and NC interactions give contribution 
to the amplitude of elastic  $\nu_{e}-e $ scattering.
The amplitude of
the elastic
$\nu_{\mu}(\nu_{\tau}) - e $ scattering is determined
only
by the NC interaction. Thus, the refraction indexes of the $\nu_{e}$
and $\nu_{\mu} ( \nu_{\tau}$) are different. Hence, when 
a neutrino wave propagates through matter, the flavor content of
the neutrino
state is changing. Under 
the condition

\begin{equation}
\Delta m^2 \cos{2\theta} = 2\sqrt{2}
\rm{G}_{\rm{F}}\rho_e E
\end{equation}
where $\rho_e $ is the electron number-density,
the combined effect of 
neutrino masses and mixing and coherent neutrino interaction in matter  
can enhance significantly the  probability of the transition of
$\nu_{e}$'s into other
states. This is so-called Mikheev-Smirnov-Wolfenstein effect (MSW). 
In the sun matter MSW effect can be important
if $ 10^{-7} \leq \Delta m^2 \leq 10^{-4} eV^{2}$.

All existing solar neutrino data can be described, if we assume
that there
is mixing of two neutrinos and the values of the solar neutrino
fluxes are given by the SSM. In such a case there are only two free
parameters: $\Delta m^{2} $ and $\sin^{2} 2 \theta $.
{}From the fit of the events rates, measured in all solar neutrino
experiments, there were found two MSW fits with large and small 
mixing angle (correspondingly, LMA and SMA)

$$ 10^{-5} < \Delta m^{2} < 10^{-4}\rm{eV}^{2}~~~ 0.8 < sin^{2}2 \theta <
1 $$

$$ 10^{-5} < \Delta m^{2} <6\cdot 10^{-6}\rm{eV}^{2}~~~4\cdot 10^{-3} <
sin^{2}2
\theta < 10^{-2}$$

The events rates measured in all solar neutrino experiments can be
also described by vacuum oscillations (VO) with

$$8\cdot 10^{-11} < \Delta m^{2} < 4 \cdot 10^{-10}\rm{eV}^{2}~~~
0.6 <\sin^{2}2 \theta <1 $$

In the high-statistics Super-Kamiokande experiment the spectrum of the
recoil electrons in the process 
$\nu + e \to \nu + e$
was measured. 
If there are no oscillations this spectrum can be predicted
in a model-independent way. This is connected with the fact that in the
Super-Kamiokande experiment only neutrinos from $^{8}B$ decay, the
spectrum
of which is determined by the weak interactions, are measured. No sizable
distortion of the spectrum was observed in this experiment.

In the Super-Kamiokande experiment the day-night asymmetry
was also measured. During night neutrinos pass through the earth
and the measurement of the day-night asymmetry allows in a model-independent 
way
to measure matter effects. No significant day-night asymmetry
was observed:

$$ \frac {N-D} {(N+D)/2} =0.034 \pm 0.022 \pm 0.013 $$

These new measurements allows one to constrain the possible values of the
neutrino oscillation parameters. From the fit of all solar neutrino data
it follows
that the most favored fit is the LMA one with

$$6\cdot 10^{-5} < \Delta m^{2} <3\cdot 10^{-4}\rm {eV}^{2}~~~ 0.8 <
\sin^{2} 2\theta < 1 $$

If solar neutrino fluxes from different sources are considered as free
parameters and it is assumed that the $\nu_{e}$ transition probability is
equal to one in this case from the analysis
of the data of different solar neutrino experiments it follows
that the 
flux of $^{8} Be$ neutrinos must be strongly suppressed.   
This consequence of the general analysis of existing solar
neutrino
data
will be checked in the future BOREXINO experiment that is planned to start
in
2002 in the underground Laboratory Gran Sasso (Italy). In this experiment 
mainly medium energy $^{8} Be $ neutrinos will be detected
through the observation of the $\nu-e$ scattering in a scintillator.

In the SNO experiment (Sudbury Neutrino Observatory, Canada)
the 
solar $\nu_{e}$'s
are detected through the observation of electrons in the  CC
reaction

\begin{equation}
\nu_e\,+\, d \to e^-\,+\, p\,+\, p
\label{rea1}
\end{equation}

In  the nearest future in the SNO
experiment the solar neutrinos will be detected
also through the
observation of the neutrons from the NC process

\begin{equation}
\nu\, +\, d \to \nu\, +\, n\,+\, p
\label{rea2}
\end{equation}

Not only $\nu_{e}$'s but also  $\nu_{\mu}$'s and $\nu_{\tau}$'s
will be detected by this method.
 The comparison of the NC and CC data will allow one to obtain a
model-independent
information on the transitions of the solar $\nu_{e}$'s into other
neutrino states.

\subsection {The atmospheric neutrino experiments }

The most compelling evidence in favor of neutrino oscillations
was 
obtained recently by the atmospheric neutrino experiments. The main
source of 
atmospheric neutrinos is the following chain of the decays

\begin{equation}
\pi \to \mu\,+\, \nu_\mu ,
\qquad
\mu \to e\,+\, \nu_e \,+\,\nu_\mu,
\label{rrea1}
\end{equation}
the pions being produced in the interaction of cosmic rays with nuclei
in the
earth's
atmosphere. At relatively small energies ($ \leq 1$ GeV )
the ratio of the muon and electron neutrinos is equal to 2. 
At higher
energies this ratio becomes larger than 2 (not 
 all muons have enough time to decay in the atmosphere).
The ratio  can be predicted, however, with the
accuracy better than 5 \%. The absolute fluxes of the electron and
muon
neutrinos are predicted at present with accuracy 25-30 \%.
The results of the atmospheric neutrino experiments are usually
presented in
the form of the double ratio $R$ of the ratio of the observed muon and
electron
events
to the ratio of the muon and electron events calculated by Monte
Carlo
method under the assumption that there are no neutrino oscillations.
In all latest atmospheric neutrino experiments it was found that
the ratio  $R$ is significantly smaller than one:

\begin{eqnarray}
R & = & 0.65 \pm 0.05 \pm 0.08 \qquad (\rm{Kamiokande})
\nonumber \\
R & = & 0.54 \pm 0.05 \pm 0.11 \qquad (\rm{IMB})
\nonumber \\
R & = & 0.61 \pm 0.15 \pm 0.05 \qquad (\rm{Soudan2})
\nonumber \\
R & = & 0.638 \pm 0.017 \pm 0.050
\qquad (\rm{Super-Kamiokande})
\end{eqnarray}

The fact that the double ratio $R$ is less than one 
is a model-independent indication in favor of the disappearance of 
 $\nu_{\mu}$ (or appearance of  $\nu_{e}$). 

Compelling evidence in
favor of
the disappearance of  $\nu_{\mu}$  was obtained 
recently by the Super-Kamiokande experiment.
In this experiment a significant zenith angle
dependence of the number of high-energy
muon events was found (the zenith angle 
$\theta$
is the
angle between the vertical direction and the neutrino momentum).
The angle $\theta$ is connected with the distance that neutrinos pass 
from the production region to the detector. Down-going
neutrinos
($\cos \theta = 1$) pass a distance of about 20 km. The distance
that up-going neutrinos ($\cos \theta = -1$)
travel is about 13000 km. 

The possible source of the zenith angle
dependence of the numbers of atmospheric neutrino events is the 
magnetic field of the earth.
However, at energies larger than 1 GeV the effect of the magnetic field of
the
earth is small and the numbers of down-going 
and up-going  $\nu_{\mu}$ ( $\nu_{e}$) must be equal.

 The Super-Kamiokande collaboration observed the 
significant up-down asymmetry of the muon events:

\begin{equation}
A_\mu = \frac {U - D} {U+D} =- 0.311 \pm 0.043 \pm 0.010
\label{frac1}
\end{equation}

Here $U$ is the total number of up-going muons 
and  $D$
is the total number of down-going muons.

For the up-down asymmetry of the electron events a value compatible
with
zero was found:

\begin{equation}
A_e = 0.036 \pm 0.067 \pm 0.02
\end{equation}

The data that was obtained by the Super-Kamiokande collaboration
can be explained by 
$\nu_{\mu}\to \nu_{\tau}$ neutrino oscillations. From the analysis of
the
data for the parameters  $\Delta m^{2}$ and 
 $\sin^{2} 2 \theta $ the following best-fit values were obtained

\begin{equation}
\Delta m^2 =  2.5\cdot 10^{-3} {\rm{eV}}^2 ,
\qquad
\sin^2 2\theta = 1
\end{equation}

The disappearance of the up-going muon neutrinos
is due to the fact that these neutrinos
travel longer distance 
than the down-going muon neutrinos and
have more time to transfer
into  $\nu_{\tau}$.

The $\nu_{\mu}$ survival probability 
depends on the ratio $L/E$ and 
is given by the expression

\begin{equation}
{\rm P}(\nu_\mu \to \nu_\mu) =1 - \frac {1}{2} \sin^2
2\theta
\left(1 - \cos\,  2.54 \Delta m^2\frac {L} {E} \right)
\label{SURVIV}
\end{equation}

At  $L/E \geq 10^{3} \rm {km/GeV} $ 
the argument of the cosine in the expression (\ref{SURVIV}) is large
and the
cosine in this expression disappears due to averaging over the 
neutrino
energies and distances. As a result at  $L/E \geq 10^{3} km/GeV$ 
for the averaged survival probability we have 
$\bar{\rm{P}}(\nu_\mu \to \nu_\mu) = 1- \frac
{1}{2} \sin^2 2\theta \simeq \frac{1}{2} $ 

The atmospheric neutrino range  $\Delta m^{2} \simeq 10 ^{-3}
eV^{2}$
will be probed the long-baseline (LBL) accelerator
neutrino experiments.
The first LBL experiment K2K have started in Japan in 1999.
The distance between the source (accelerator)
and the 
detector (Super-Kamiokande) is about 250 km.
Two other LBL experiments are under preparation.
In the MINOS experiment neutrinos produced
from the accelerator at Fermilab (USA) will be detected by the detector
in the Soudan mine (the distance is about 730 km). In another LBL
experiment
neutrinos produced from the accelerator at CERN (Geneva) will be
detected
by the detector at the underground Laboratory Gran Sasso (Italy) (the
distance is also about 730 km). In the accelerator experiments initial
neutrinos
are
mainly $\nu_ {\mu}$ with a small admixture of  $\nu_ {e}$. In 
the CERN-Gran Sasso experiment
appearance of  $\nu _{\tau}$ will be searched for.

\subsection {The LSND experiment}

Some indications in favor of  $\bar\nu_{\mu}\to \bar\nu_{e}$ oscillations
were obtained also in the short-baseline experiment that was done
at the Los Alamos linear accelerator (USA).
In this experiment a beam of pions produced by 800 MeV
protons hits a copper target. In this target the $\pi^+$-mesons come
to 
rest and decay ($\pi^{+} \to \mu^{+} + \nu_{\mu}$).
The produced muons also come to rest in the target
and decay ( $\mu^{+} \to e^{+} + \bar \nu_{\mu} + \nu_{e}$).
Thus, in decays of the $\pi^{+}$'s
and $\mu^{+}$'s muon neutrinos  $\nu_{\mu}$, muon antineutrinos $\bar
\nu_{\mu}$
and electron neutrinos $\nu_{e}$ are produced. There is no
electron antineutrinos $\bar\nu_{e}$ from these decays .
Let us notice that  $\bar\nu_{e}$'s are produced in the decay chain  
that starts with $\pi^{-}$'s. However, practically all $\pi^{-}$'s are
captured by nuclei in the target and have no time to decay.

In the LSND  neutrino detector at a distance of about 30 m
from the target, the electron antineutrinos $\bar \nu_{e}$'s were
searched for
through the observation
of the classical process

\begin{equation}
\overline{\nu}_e + p \to e^+ + n
\end{equation}

In the interval of the positron energies $30 < E < 60 $ MeV  
it was observed in the LSND experiment 
 $87.9 \pm 22.4 \pm 6.0 $
events.

The observed signal can be explained by  
 $\bar \nu_ {\mu}\to \bar \nu_ {e}$ oscillations.
If we take into account the results of the other short-baseline
experiments in
which neutrino oscillations were not found, from the LSND
experiment the following ranges of the oscillation parameters can be
found

\begin{equation}
0.2 \leq \Delta m^2 \leq 1 \rm{eV}^2 \qquad
2 \cdot 10^{-3} \leq \sin^2 2\theta \leq 4 \cdot 10^{-2}
\end{equation}

The indications in favor of 
 $\nu_{\mu}\to \nu_{e}$  oscillations, obtained in the LSND experiment,
will
be checked by the BOONE experiment 
(Fermilab, USA) that will  start in 2002.

\section {Neutrinoless double $\beta$-decay}

We have discussed in the previous sections 
neutrino oscillation experiments that 
allow to obtain information on a very small 
neutrino mass squared differences.
Important information on the neutrino masses 
and {\it the nature of massive neutrinos}
can be obtained from experiments on the investigation
of neutrinoless 
 double $\beta$-decay
\begin{equation}
(A,Z) \to (A,Z+2) + e^- + e^-
\label{AZ2}
\end{equation}

Here $(A, Z)$ is some even-even nucleus. In the experiments   
neutrinoless double  $\beta$-decay of $ ^{76}\rm{Ge} $,
$ ^{136}\rm{Xe} $, $ ^{130}\rm{Te} $, $ ^{100}\rm{Mo} $ and other nuclei
are searched
for. 
The process (\ref{AZ2}) is allowed, if the total lepton number $L$ is
not
conserved, i.e. if massive neutrinos are Majorana particles.

In the framework of the standard CC weak interaction with Majorana
neutrino mixing neutrinoless double $\beta$-decay is second
order in
the
Fermi constant $G_{F}$  process with a virtual neutrino. 
The matrix element of the process is proportional to the
effective Majorana mass

\begin{equation}
\langle m \rangle = \sum_i U_{e i}^2 m_i
\end{equation}
where $m_{i}$
is the neutrino mass.  

There are many experiments in which neutrinoless
 double $\beta$-decay of different nuclei are searched for. No positive
indications in favor
of such decay were found up to now. A very stringent lower bound
on the life-time
was obtained in 
the Heidelberg-Moscow experiment in which the neutrinoless double
$\beta$-decay of the ${}^{76} $Ge  was search for:
\begin{equation}
T_{1/2} >  1.6 \times 10^{25} \, \rm{years}
\end{equation}

The upper bound of the effective Majorana mass that can be obtained
from this result depends on the calculation of nuclear matrix
elements. Using different calculations one can find

\begin{equation}
|\langle m \rangle| > (0.3 - 0.9)\,\, {\rm{eV}}
\end{equation}

In the next generation experiments on the search for 
 neutrinoless double $\beta$-decay the sensitivity
$ |\langle m \rangle| \simeq  10^{-1}$ eV
will be achieved (NEMO3, Heidelberg-Moscow,IGEX). The possibility of 
the
experiments,
in which the sensitivity $ |\langle m \rangle| \simeq  10^{-2}$ eV will be
reached,
is under investigation.

\section {Neutrino masses from experiments on the measurement of
the $ \beta $-spectrum of tritium }

The first method of measuring neutrino mass was proposed in the
classical
paper by Fermi on the $ \beta $-decay.
The method consists in the precise measurement of the end-point
part
of the $ \beta $-spectrum, the part of the spectrum that is most sensitive
to
the small neutrino mass.

Usually, for the determination of neutrino mass  by this method the 
$ \beta $-spectrum 
of the decay of the tritium

\begin{equation}
{}^{3}\rm{H} \to {}^{3}{\rm{He}} + e^- + \bar{\nu}_e
\label{two76}
\end{equation}
is investigated.
The  $ \beta $-spectrum of this decay 
is determined by the phase-space factor 

\begin{equation}
\frac {dN}{dT} = C\, p E (Q - T)
\sqrt{(Q - T)^2 - m^2_\nu} F(E)
\label{two75}
\end{equation}

Here $p$ and $E$ are the electron momentum and energy, 
respectively, $T =E-m_{e}$ is
the electron
kinetic energy, $ Q \simeq 18.6$ keV is
the energy release,
$ C= const $,
$F( E) $ is the known function that describes the Coulomb
interaction of
the final particles and $ m_{\nu } $ is the mass of the $\nu_{e}$.
If the neutrino mass is equal to zero,  $T_{max}= Q$. For nonzero
neutrino mass
$T_{max}= Q - m_{\nu}$. Thus, for nonzero neutrino mass
at the end-point part of the electron spectrum
 the deficit of the
events (with respect to the number of the events expected for
 $ m_{\nu }=0 $) must be observed. 

At the moment no positive indications in favor of nonzero
neutrino mass were obtained from the 
${}^3$H  experiments. For the upper bound of the neutrino mass it was
found

\begin{eqnarray}
m_\nu \leq 2.5~~ \rm{eV}~~~~~(\rm{Troitsk})
\nonumber \\
m_\nu \leq 2.2~~ \rm{eV}~~~~~(\rm{Mainz})
\end{eqnarray}
In future experiments on the measurement of the end-point part of the
spectrum of 
$\beta$-decay of $^3H$ the sensitivity $m_{\nu} \simeq 0.5$ eV is planned
to be
achieved.

\section {Conclusion}

The neutrinos 
play very a important role in particle physics and astrophysics.
They have enormous penetration
properties
and they give us a unique possibility to investigate the
internal 
structure of the nucleon, the internal invisible region of the sun 
where solar energy is produced etc. 

The neutrinos are exceptional particles as for their
internal
properties.  The neutrino masses are many orders of magnitude smaller
than the
masses of their family partners (electron, muon, tau). Because
of the smallness of the neutrino masses new physical phenomenon, 
{\it neutrino oscillations}, the periodical transitions between
different
flavor
neutrinos in the vacuum or in matter, becomes possible.  
The evidence for this phenomenon, that was predicted many years ago, was
obtained recently by the Super-Kamiokande collaboration in Japan.
The investigation of the neutrino oscillations that is going on all
over the
world is a new field of research in particle physics and astrophysics.

The investigation of the neutrino oscillations, 
neutrinoless double $\beta$-decay, $\beta$-spectrum of
$^3 H$-decay and other effects
will allow us to
obtain important information on the neutrino masses, 
element of the
neutrino
mixing matrix and the nature of massive neutrinos (Dirac or Majorana?). 

The exceptional smallness of the neutrino masses requires a special
explanation. There is a general belief that   
small neutrino masses are generated by new interactions
beyond the Standard Model.
One of the plausible explanation of the small neutrino masses is connected
with a violation of the lepton number at a mass scale 
that is much larger than the scale of the violation of electroweak  
symmetry $M_{EW} \simeq 10^ {2}$ GeV  that determine 
masses of
the leptons, 
quarks and $W^{\pm}$, $Z^{0}$ bosons.
If this explanation is correct the massive neutrinos are
truly neutral Majorana particles. All
other fundamental fermions (leptons and quarks) are charged Dirac
particles.

It is a pleasure for me to thank Tommy Ohlsson for careful reading of the
paper and useful discussions and Michael Ratz for the help in the
preparation of the figures.


\begin{thebibliography}{100}
\bibitem{}
J.N. Bahcall,
{\it{Neutrino Physics and Astrophysics}}, Cambridge University
Press, 1989.

\bibitem{}
V. Barger, {\it Neutrino masses and Mixing at the Millenium},
hep-ph/0003212.



\bibitem{Bilenky-Pontecorvo78}
S.M. Bilenky and B.~Pontecorvo,             
Phys. Rep. {\bf{41}}, 225 (1978).

\bibitem{}
S.M. Bilenky and S.T. Petcov.
Rev. Mod, Phys, 59, 671 (1989).



\bibitem{BGG}
S.M.~Bilenky, C.~Giunti and W.~Grimus,
Progr. in Part. and Nucl. Phys. {\bf{43}}, 1 (1999).
             

\bibitem{}   
F.~Boehm and P.~Vogel,
{\it{Physics of Massive Neutrinos}}, Cambridge University Press,
Cambridge,
  1994.

\bibitem{}
J.~Brunner,
Fortsh. Phys. {\bf{45}}, 343 (1997)




\bibitem{} 

W.~Haxton, B.~Holstein,
{\it{Neutrino Physics}}, Am. J. Phys. {\bf{68}},15 (2000).
\bibitem{}
B. Kayser, F. Gibrat-Debu and Perrier, 
{\it The Physics of Massive Neutrinos}, World Scientific Singapore, 1989.



\bibitem{Kim93}
C.W. Kim and A.~Pevsner,
{\it{Neutrinos in Physics and Astrophysics}}, Contemporary Concepts
in
  Physics, Vol. 8, Harwood Academic Press, Chur, Switzerland, 1993.


\bibitem{Mohapatra-Pal91}
R.N. Mohapatra and P.B. Pal,  
{\it{Massive Neutrinos in Physics and Astrophysics}}, World
Scientific, Singapore, 1998.

\bibitem{}
{\it{Neutrino Physics}}, 2th edition, editor K.~ Winter, Cambridge
University
Press, Cambridge, 2000.


\bibitem{Oberauer-vonFeilitzsch92}
L.~Oberauer and F.~von Feilitzsch,
Rept. Prog. Phys. {\bf{55}}, 1093 (1992).

\bibitem{}
A. Pais
{\it Inward bound: of Matter and Forces in the Physical World},
Oxford University Press, 1986.


\bibitem{}
D.H. Perkins,
{\it{Introduction to high Energy Physics}}, 4th edition, Cambridge
University
Press, Cambridge, 2000.

\bibitem{}
P. Renton
{\it Electroweak Interactions - An Introduction to the Physics of Quarks
and
Leptons}, Cambridge, Cambridge University Press, 1990.




\bibitem{Zuber98-NEUTRINOS}
K.~Zuber,
Phys. Rep. {\bf{305}}, 295 (1998).






\end{thebibliography}
\end{document}